\documentclass[11pt,doublecolumn,twosided]{IEEEtran}%
\usepackage{graphics,graphicx,psfrag,epsfig,color,rotating}
\usepackage{amssymb,amscd,amsmath,amsfonts}
\usepackage{dsfont,extarrows}
\usepackage{lscape}
\usepackage{array,multirow}
\usepackage{url}
\usepackage[english]{babel}

\newtheorem{lemma}{Lemma}
\newtheorem{theorem}{Theorem}
\newtheorem{definition}[theorem]{Definition}
\newtheorem{proposition}[theorem]{Proposition}

\newcommand{\mb}{\mathbf}

\newcommand{\bs}{\mathbf}

\newcommand{\defn}{\stackrel{\triangle}{=} }

\renewcommand{\d}{\mathrm{d}}
\newcommand{\Id}{{\bf I}}
\newcommand{\im}{{\mathrm{Im}}}
\newcommand{\tr}{{\mathrm{Trace}}}
\newcommand{\diag}{{\mathrm{diag}}}
\newcommand{\Rplus}{{\mathds{R}^+}}
\newcommand{\Ker}{{\mathrm{Ker}}}
\newcommand{\rank}{{\mathrm{rank}}}
\newcommand{\Span}{{\mathrm{Span}}}

\def\ind{{\mathds{1}}}
\definecolor{orange}{RGB}{255,128,0}

\definecolor{dgreen}{RGB}{0,151,0}

\begin{document}
\title{From Spectrum Pooling to Space Pooling: Opportunistic Interference Alignment in MIMO Cognitive Networks}
\author{S.M.~Perlaza,
         N.~Fawaz,
         S.~Lasaulce,
         ~and~M.~Debbah,
\thanks{S.M. Perlaza is with Orange Labs, France Telecom R\&D. $38-40$ rue du G\'en\'eral Leclerc, 92794, Issy les Moulineaux, cedex $9$. France. (samir.medinaperlaza@orange-ftgroup.com)}%
\thanks{N.~Fawaz is with the Research Laboratory of Electronics, Massachusetts Institute of Technology (MIT). Cambridge, Massachusetts, MA-02139, USA.    (nfawaz@mit.edu). 
}%
\thanks{S. Lasaulce is with LSS (CNRS-SUPELEC-Paris Sud). $3$ rue Joliot-Curie,
 $91192$, Gif-sur-Yvette, cedex. France. (lasaulce@lss.supelec.fr) }%
\thanks{M. Debbah is with Alcatel-Lucent Chair in Flexible Radio at SUPELEC. $3$ rue Joliot-Curie,
$91192$, Gif-sur-Yvette, cedex. France. (merouane.debbah@supelec.fr)}%
\thanks{The material in this paper has been presented in part at PIMRC 2008. September 2008, Cannes, France \cite{IA-Perlaza-08b}.}
}

\maketitle
\begin{abstract}
\boldmath
We describe a non-cooperative interference alignment (IA) technique which allows an opportunistic
 multiple input multiple output (MIMO) link (secondary) to harmlessly coexist
 with another MIMO link (primary) in the same frequency band. Assuming perfect channel knowledge at the primary receiver and transmitter, capacity is achieved by  transmiting along the spatial directions (SD) associated with the singular values of its channel matrix using a water-filling power allocation (PA) scheme. Often, power limitations lead the primary transmitter to leave some of its SD unused. Here, it is shown that the opportunistic link can transmit its own data if it is possible to align the interference produced on the primary link with such unused SDs. We provide both a processing scheme to perform IA and a PA scheme which maximizes the transmission rate of the opportunistic link. The asymptotes of the achievable transmission rates of the opportunistic link are obtained in the regime of large numbers of antennas. Using this result, it is shown that depending on the signal-to-noise ratio and the number of transmit and receive antennas of the primary and opportunistic links, both systems can achieve transmission rates of the same order.
\end{abstract}

\section{Introduction}\label{SecIntroduction}

The concept of cognitive radio is well-known by now. The main idea
is to let a class of radio devices, called secondary systems,
opportunistically access certain portions of spectrum left unused by
other radio devices, called primary systems, at a given time or
geographical area  \cite{Haykin-05}. These pieces of unused
spectrum, known as white-spaces, appear mainly when either
transmissions in the primary network are sporadic, i.e., there are
periods over which no transmission takes place, or there is no
network infrastructure for the primary system in a given
area, for instance, when there is no primary network coverage in
a certain region. In the case of dense networks, a white-space
might be a rare and short-lasting event. As a matter of fact, the
idea of cognitive radio as presented in \cite{Haykin-05} (i.e.,
spectrum pooling), depends on the existence of such white-spaces
\cite{Haddad-08}. In the absence of those spectrum holes,
secondary systems are unable to transmit  without producing additional interference on
the primary systems. One solution to this situation has been
provided recently under the name of interference alignment (IA)
\cite{IA-Jafar-08a}. Basically, IA refers to the construction of
signals such that the resulting interference signal lies in a
subspace orthogonal to the one spanned by the signal of interest at
each receiver \cite{IA-Jafar-07a}. The IA concept was introduced separetely and
almost simultaneously by several authors \cite{IA-Khandani-06,
IA-Khandani-08a, IA-Jafar-08a, IA-Shamai-07a}. Recently, IA has become an important tool to study the interference
channel, namely its degrees of freedom \cite{IA-Jafar-07a,
IA-Jafar-08a, IA-Jafar-08c}. The feasibility and implementation
issues of IA regarding mainly the required channel state information
(CSI) has been also extensively studied \cite{Heath-09, Bolcskei-09,
IA-Jafar-08b, IA-Tresh-09}.

\noindent
In this paper we study an IA scheme named opportunistic IA (OIA) \cite{IA-Perlaza-08b}. The idea behind OIA can be briefly described as follows. The primary link is modeled by a single-user MIMO channel since it must operate free of any additional interference produced by secondary systems.
Then, assuming perfect CSI at both transmitter and receiver ends, capacity is achieved by implementing
a water-filling power allocation (PA) scheme \cite{Telatar-99} over the spatial
directions associated with the singular values of its channel
transfer matrix. Interestingly, even if the primary transmitters
maximize their transmission rates, power limitations generally lead
them to leave some of their spatial directions (SD) unused. The unused SD can therefore be reused by another system operating in the same frequency band. Indeed, an opportunistic transmitter can send its own data to its respective receiver by processing its signal in such a way that the interference produced on the primary link impairs only the unused SDs. Hence, these spatial resources can be very useful for a secondary system when the available spectral resources are fully exploited over a certain period in a geographical area.  The idea of OIA, as described above, was first introduced in \cite{IA-Perlaza-08b} considering a very restrictive scenario, e.g., both primary and secondary devices have the same number of antennas and same power budget. In this paper, we consider a more general framework where devices have different number of antennas, different power budgets and no conditions are impossed over the channel transfer matrices (In \cite{IA-Perlaza-08b}, full rank condition was impossed over certain matrices).

\noindent
The rest of this paper is structured as follows. First, the system model, which
consists of an interference channel with MIMO links, is
introduced in Sec. \ref{SecSystemModel}. Then, our aim in Sec.
\ref{SecIAStrategy} is twofold. First, an analysis
of the feasibility of the OIA scheme is provided. For this
purpose, the existence of transmit
opportunities (SD left unused
 by the primary system) is studied. The average number of transmit
opportunities is expressed as a function of the number of antennas
at both the primary and secondary terminals. Second, the proposed interference alignment technique and power
allocation (PA) policy at the secondary transmitter are
described. In Sec. \ref{SecAsymptoticAnalaysis}, tools from random matrix theory for large systems are
used to analyze the achievable transmission rate of the
opportunistic transmitter when no optimization is performed over its
input covariance matrix. We illustrate our theoretical results by
simulations in Sec. \ref{SecSimulations}. Therein, it is shown 
 that our approach allows the secondary link to achieve
transmission rates of the same order as those of the primary link.
Finally, in Sec. \ref{SecConclusions} we state our conclusions and
provide possible extensions of this work.

\section{System Model}\label{SecSystemModel}

\emph{Notations}. In the sequel, matrices and vectors are
respectively denoted by boldface upper case symbols and boldface
lower case symbols. An $N \times K$ matrix with ones on its main
diagonal and zeros on its off-diagonal entries is denoted by
$\Id_{N\times K}$, while the identity matrix of size $N$ is simply
denoted by $\Id_N$. An $N \times K$ matrix with zeros in all its
entries (null matrix) is denoted by $\bs{0}_{N \times K}$. Matrices
$\bs{X}^T$ and $\bs{X}^H$ are the transpose and Hermitian transpose
of matrix $\bs{X}$, respectively. The determinant of matrix $\bs{X}$
is denoted by $|\bs{X}|$. The expectation operator is denoted by
$\mathbb{E}\left[.\right]$. The indicator function associated with a
given set $\mathcal{A}$ is denoted by $\mathds{1}_{\mathcal{A}}(.)$,
and defined by $\mathds{1}_{\mathcal{A}}(x) = 1$ (resp. $0$) if $x
\in \mathcal{A}$ (resp. $x \notin \mathcal{A}$). The Heaviside step
function and the Dirac delta function are respectively denoted by
$\mu(\cdot)$ and $\delta(\cdot)$. The symbols $\mathds{N}$,
$\mathds{R}$,  and $\mathds{C}$ denote the sets of non-negative
integers, real numbers, and complex numbers, respectively. The
subsets $\left[0,+\infty \right[$ and $\left]-\infty,0\right]$ are
denoted by $\mathds{R}^+$ and $\mathds{R}^-$, respectively. The
operator $\left(x \right)^+$ with $x \in \mathds{R}$ is equivalent
to the operation $\max\left( 0, x \right)$. Let  $\bs{A}$ be an $n
\times n$ square matrix with real eigenvalues $\lambda_{A,1},
\ldots, \lambda_{A,n}$. We define the empirical eigenvalue
distribution of $\bs{A}$ by $F_{A}^{(n)}(\cdot) \triangleq
\frac{1}{n}\sum_{i=1}^{n}\mu(\lambda-\lambda_{A,i})$,
and, when it exists, we denote $f_{A}^{(n)}(\lambda)$ the associated
eigenvalue probability density function, where $F_{\bs{A}}(\cdot)$
and $f_{\bs{A}}(\cdot)$ are respectively the associated limiting
eigenvalue distribution and probability density function when
$n\rightarrow + \infty$.

\begin{figure}[!ht]
\begin{center}
\psfrag{A1}{1}
\psfrag{A2}{2}
\psfrag{A3}{$M_1$}
\psfrag{B1}{1}
\psfrag{B2}{2}
\psfrag{B3}{$N_1$}
\psfrag{C1}{1}
\psfrag{C2}{2}
\psfrag{C3}{$M_2$}
\psfrag{D1}{1}
\psfrag{D2}{2}
\psfrag{D3}{$N_2$}
\psfrag{H11}{$H_{11}$}
\psfrag{H21}{$H_{21}$}
\psfrag{H12}{$H_{12}$}
\psfrag{H22}{$H_{22}$}
\psfrag{TX1}{$Tx_1$}
\psfrag{TX2}{$Tx_2$}
\psfrag{RX1}{$Rx_1$}
\psfrag{RX2}{$Rx_2$}
\psfrag{Primary}{\hspace{-7mm} Primary System}
\psfrag{Secondary}{\hspace{-6mm} Secondary System}
\includegraphics[width=.9\linewidth]{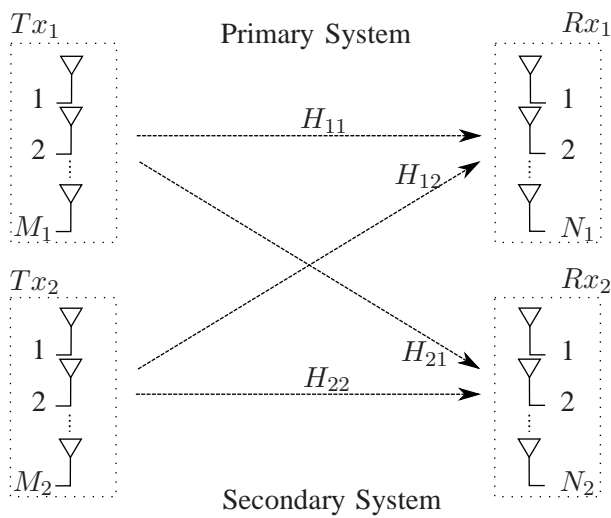}
\caption{\label{FigMIMOIC} Two-user MIMO interference channel.}
\end{center}
\end{figure}

\noindent
We consider two  unidirectional links  simultaneously operating in
the same frequency band and producing mutual interference as shown in Fig. \ref{FigMIMOIC}. The first transmitter-receiver pair
$(\mathrm{Tx}_1, \mathrm{Rx}_1)$ is the primary link. The pair
$(\mathrm{Tx}_2, \mathrm{Rx}_2)$ is an opportunistic link subject to
the strict constraint that the primary link must transmit at a rate
equivalent to its single-user capacity. Denote by $N_i$ and
$M_i$, with $i = 1$ (resp. $i = 2$), the number of antennas at the
primary (resp. secondary) receiver and transmitter, respectively.
Each transmitter sends independent
messages only to its respective receiver and no cooperation
between them is allowed, i.e., there is no message exchange
between transmitters. This scenario is known as the MIMO
interference channel (IC) \cite{Jafar-04, Gans-06} with private
messages. A private message is a message from a given source
to a given destination: only one destination node is able to decode
it. Indeed, we do not consider the case of common messages which
would be generated by a given source in order to be decoded by
several destination nodes.

\noindent
In this paper, we assume the channel transfer matrices between different nodes to be fixed over the
whole duration of the transmission. The channel transfer matrix from transmitter $j  \in
\left\lbrace 1,2 \right\rbrace$ to receiver $i \in \left\lbrace 1,
2\right\rbrace$ is an $N_i \times M_j$ matrix denoted by
$\bs{H}_{ij}$ which corresponds to the realization of a random
matrix with independent and identically distributed (i.i.d.)
complex Gaussian circularly symmetric entries with zero mean
and variance $\frac{1}{M_j}$, which implies
\begin{equation}\label{EqHVariance}
\forall (i,j) \in \lbrace 1,2 \rbrace^2, \quad \mathrm{Trace}\left(\mathds{E}\left[\bs{H}_{ij} \: \bs{H}_{ij}^H \right]\right) = N_i.
\end{equation}
The  $L_i$  symbols transmitter $i$ is able to simultaneously transmit, denoted by $ s_{i,1}, \ldots, s_{i,L_i}$, are
represented by the vector $\bs{s}_i = \left( s_{i,1}, \ldots,
s_{i,L_i}\right)^T$. We assume that $\forall i \in \lbrace 1,2\rbrace$ symbols $s_{i,1}, \ldots, s_{i,L_i}$ are i.i.d. zero-mean circularly-symmetric complex Gaussian variables. In our model, transmitter $i$ processes its
symbols using a matrix $\bs{V}_i$ to construct its transmitted
signal $\bs{V}_i \bs{s}_i$. Therefore, the matrix $\bs{V}_i$ is
called pre-processing matrix. Following a matrix notation, the
primary and secondary received signals, represented by the
 $N_i\times 1$ column-vectors $\bs{r}_i$, with $i \in \lbrace 1,2 \rbrace$, can be written as
\begin{equation}\label{EqReceivedSignal}
\left( \begin{array}{c} \bs{r}_1 \\
\bs{r}_2\end{array}\right) = \left( \begin{array}{cc}
\bs{H}_{11} & \bs{H}_{12} \\ \bs{H}_{21}
& \bs{H}_{22} \end{array}\right) \left(\begin{array}{c}
\bs{V}_1  \bs{s}_1 \\ \bs{V}_2
\bs{s}_2 \end{array}\right) + \left( \begin{array}{c}
\bs{n}_1 \\ \bs{n}_2 \end{array}\right),
\end{equation}
where $\bs{n}_i$ is an $N_i$-dimensional vector representing noise
effects at receiver $i \in \lbrace1,2\rbrace$ with entries modeled by an additive white
Gaussian noise (AWGN) process with zero mean and variance
$\sigma^2_i$, i.e.,$\forall i \in \lbrace 1,2\rbrace$,  $\mathbb{E}\left[\bs{n}_i \bs{n}_i^H \right] =
 \sigma^2_i\Id_{N_i}$. At transmitter $i \in \lbrace 1,2\rbrace$, the $L_i \times L_i$ power allocation matrix $\bs{P}_i $ is defined by the input covariance
matrix $\bs{P}_i = \mathbb{E}\left[\bs{s}_i \bs{s}_i^H \right]$. Note that symbols $s_{i,1} \ldots, s_{i,L_i}$, $\forall i \in \lbrace 1,2\rbrace$ are mutually independent and zero-mean, thus, the PA matrices can be written as diagonal matrices, i.e., $\bs{P}_i = \diag\left(p_{i,1}, \ldots, p_{ i,L_i}\right)$. Choosing $\mathbf{P}_i$ therefore means selecting a given PA policy. The power constraints on the transmitted signals $\bs{V}_i \bs{s}_i$ can be written as
\begin{equation}\label{EqPowerConstraints}
\forall i \in \left\lbrace 1, 2 \right\rbrace, \quad \text{Trace}\left( \bs{V}_i \bs{P}_{i} \bs{V}_i^H \right) \leqslant M_i \: p_{i,\max}.
\end{equation}

\noindent
Note that assuming that the i.i.d. entries of matrices
 $\bs{H}_{ij}$, for all $(i,j) \in \left\lbrace 1,2 \right\rbrace^2$, are
  Gaussian random variables with zero mean and variance $\frac{1}{M_j}$, together with the power constraints in (\ref{EqPowerConstraints}), is equivalent to considering
 a system where the entries of matrices $\bs{H}_{ij}$ for all $(i,j) \in \left\lbrace 1,2 \right\rbrace^2$ are
  Gaussian random variables with zero mean and unit variance,  and the transmitted signal $\bs{V}_i \bs{s}_i$
 are constrained by a finite transmit power $p_{i,\max}$. Nonetheless, the second
  convention allows us to
  increase the dimension of the system (number of antennas) while maintaining
  the same average received signal to noise ratio (SNR) level
   $\frac{p_{i,\max}}{\sigma_i^2}$, $\forall i \in \left\lbrace 1,2 \right\rbrace$. Moreover, most of the tools from random matrix theory used in the asymptotic analysis of the achievable data rate of the opportunistic link in Sec. \ref{SecAsymptoticAnalaysis},
    require the variance of the entries of  channel matrices to be normalized by its size. That is the reason why the normalized model, i.e.,  channel transfer matrices and power constraints respectively satisfying (\ref{EqHVariance}) and (\ref{EqPowerConstraints}), was adopted.

\noindent
At receiver $i\in\left\lbrace 1,2 \right\rbrace$, the signal $\bs{r}_i$
is processed using an $N_i \times N_i$ matrix $\bs{D}_i$ to form the $N_i$-dimensional
 vector $\bs{y}_i = \bs{D}_i \bs{r}_i$. All along this paper, we refer to $\bs{D}_i$ as
  the post-processing matrix at receiver $i$. Regarding channel knowledge
assumptions at the different nodes, we assume that the primary
terminals (transmitter and receiver) have perfect knowledge of the
matrix $\bs{H}_{11}$ while the secondary terminals have perfect
knowledge of all channel transfer matrices $\bs{H}_{ij}$, $\forall
(i,j)\in \left\lbrace 1, 2 \right\rbrace^2$. One might ask whether
this setup is highly demanding in terms of information assumptions.
In fact, there are several technical arguments making this setup
relatively realistic: (a) in some contexts channel reciprocity can
be exploited to acquire CSI at the transmitters; (b) feedback
channels are often available in wireless communications
\cite{Bolcskei-09}, and (c) learning mechanisms \cite{IA-Jafar-08b}
can be exploited to iteratively learn the required CSI. In any case,
the perfect information assumptions provide us with an upper bound
on the achievable transmission rate for the secondary link.

\section{Interference Alignment Strategy} \label{SecIAStrategy}

In this section, we describe how both links introduced in Sec.
\ref{SecSystemModel} can simultaneously operate under the constraint
that no additional interference is generated by the opportunistic
transmitter on the primary receiver. First, we revisit the
transmitting scheme implemented by the primary system
\cite{Telatar-99}, then we present the concept of transmit
opportunity, and finally we introduce the proposed opportunistic IA
technique.

\subsection{Primary Link Performance}\label{SecIAPrimaryPerformance}

According to our initial assumptions (Sec. \ref{SecSystemModel}) the
primary link must operate at its highest transmission rate in the
absence of interference. Hence, following the results in
\cite{Telatar-99, Telatar-95} and using our own notation, the optimal pre-processing and
post-processing schemes for the primary link are given by the
following theorem.

\noindent
\begin{theorem}\label{ThTelatar95} \emph{Let $\bs{H}_{11} = \bs{U}_{H_{11}} \bs{\Lambda}_{H_{11}} \bs{V}^H_{H_{11}}$ be a singular value decomposition (SVD) of the $N_1 \times M_1$ channel transfer matrix $\bs{H}_{11}$, with $\bs{U}_{H_{11}}$ and $\bs{V}_{H_{11}}$, two unitary matrices with dimension $N_1\times N_1$ and $M_1 \times M_1$, respectively,  and $\bs{\Lambda}_{H_{11}}$ an $N_1 \times M_1$ matrix with main diagonal $\left(\lambda_{H_{11},1}, \ldots, \lambda_{H_{11},\min(N_1,M_1)}\right)$ and zeros on its off-diagonal. The primary link achieves capacity by choosing $\bs{V}_1 =
\bs{V}_{H_{11}}$, $\bs{D}_1 = \bs{U}_{H_{11}}^H$, $\bs{P}_1 =
\diag{\left(p_{1,1}, \ldots, p_{1, M_1}\right)}$, where
\begin{equation}\label{EqWaterfillingPrimary}
\forall n \in \left\lbrace 1,\ldots, M_1 \right\rbrace, \quad
p_{1,n} = \left(\beta - \frac{\sigma^2_1}{\lambda_{H_{11}^H H_{11},n}}\right)^+,
\end{equation}
with, $\bs{\Lambda}_{H_{11}^H H_{11}}= \bs{\Lambda}_{H_{11}}^H
\bs{\Lambda}_{H_{11}} = \diag\left(\lambda_{H_{11}^H H_{11},1},
\ldots, \lambda_{H_{11}^H H_{11},M_1}\right)$ and the constant
$\beta$ (water-level) is set to saturate the power constraint
(\ref{EqPowerConstraints}).}
\end{theorem}

\noindent
Let $N \triangleq \min(N_1,M_1)$. When implementing its capacity-achieving transmission scheme, the primary transmitter allocates its transmit power over an equivalent channel $\bs{D}_1 \bs{H}_{11} \bs{V}_1 = \bs{\Lambda}_{H_{11}}$ which consists of at most $\rank(\bs{H}_{11}^H\bs{H}_{11}) \leq N$ parallel sub-channels with non-zero channel gains $\lambda_{H_{11}^H H_{11},n}$, respectively. These non-zero channel gains to which we refer as transmit dimensions, correspond to the non-zero eigenvalues of matrix $\bs{H}_{11}^H\bs{H}_{11}$. The transmit dimension $n \in \lbrace 1, \ldots, M_1 \rbrace$ is said to be used by the primary transmitter if $p_{1,n} > 0$. Interestingly, (\ref{EqWaterfillingPrimary}) shows that some of the transmit dimensions can be left unused. Let $m_1 \in \lbrace 1, \ldots, M_1 \rbrace$ denote the number of transmit dimensions used by the primary user:
\begin{equation}\label{eq:PrimaryResDim}
\begin{split}
m_1 & \triangleq \sum_{n=1}^{M_1} \ind_{ ]0,M_1 p_{\mathrm{1,\max}} ] }(p_{1,n})\\
		& =  \sum_{n = 1}^{M_1} \mathds{1}_{\left] \frac{\sigma^2_1}{\beta},+\infty\right[}  (\lambda_{H_{11}^HH_{11},n}).
\end{split}
\end{equation}
As $p_{\mathrm{1,\max}} >0$, the primary link transmits at least over dimension $n^* = \displaystyle\arg\max_{m \in \lbrace 1, \ldots, \min(N_1,M_1)\rbrace}\left\lbrace \lambda_{H_{11}^HH_{11},m}\right\rbrace$ regardless of its SNR, and moreover, there exist at most $N$ transmit dimensions, thus
\begin{equation}\label{eq:xi1Ineq}
1 \leq m_1 \leq \rank(\bs{H}_{11}^H\bs{H}_{11}) \leq N.
\end{equation}
In the following section, we show how those unused dimensions of the primary system can be seen by the secondary system as opportunities to transmit.

\subsection{Transmit Opportunities}\label{SecTO}

Once the PA matrix is set up following Th. \ref{ThTelatar95}, the primary equivalent channel $\bs{D}_1 \bs{H}_{11} \bs{V}_1 \bs{P}_1^{1/2}= \bs{\Lambda}_{H_{11}} \bs{P}_1^{1/2} $ is an $N_1 \times M_1$ diagonal matrix whose main diagonal contains $m_1$ non-zero entries and $N-m_1$ zero entries. This equivalent channel transforms the set of $m_1$ used and $M_1-m_1$ unused transmit dimensions into a set of $m_1$ receive dimensions containing a noisy version of the primary signal, and a set of $N_1-m_1$ unused receive dimensions containing no primary signal. The $m_1$ useful dimensions are called primary reserved dimensions, while the remaining $N_1-m_1$ dimensions are named secondary transmit opportunities (TO). The IA strategy, described in Section \ref{SecV2}, allows the secondary user to exploit these $N_1-m_1$ receive dimensions left unused by the primary link, while avoiding to interfere with the $m_1$ receive dimensions used by the primary link.

\noindent
\begin{definition}[Transmit Opportunities]\label{DefTO}\emph{
Let $\lambda_{H_{11}^HH_{11},1},
\ldots \lambda_{H_{11}^HH_{11},M_1}$ be the
eigenvalues of matrix $\bs{H}_{11}^H\bs{H}_{11}$  and $\beta$ be the water-level in (Th. \ref{ThTelatar95}). Let $m_1$, as defined in (\ref{eq:PrimaryResDim}), be the number of primary reserved dimensions.
Then the number of transmit opportunities $S$ available to the opportunistic terminal is given by
\begin{equation}\label{EqSumOfS}
    S \triangleq N_1 - m_1 = N_1 - \sum_{n = 1}^{M_1} \mathds{1}_{\left] \frac{\sigma^2_1}{\beta},+\infty\right[}  (\lambda_{H_{11}^HH_{11},n}).
\end{equation}
}
\end{definition}
Note that in this definition it is implicitly assumed
that the number of TOs is constant over a duration
equal to the channel coherence time.

\noindent
Combining (\ref{eq:xi1Ineq}) and (\ref{EqSumOfS}) yields the bounds on the number of transmit opportunities
\begin{equation}\label{EqSBounds}
 N_1-N \leq S \leq N_1 -1.
\end{equation}

\noindent
A natural question arises as to whether the number of TOs is sufficiently high for the
secondary link to achieve a significant transmission rate. In order to provide an element of response to this question, a method to find an approximation of the number of TOs per primary transmit antenna, $S_{\infty}$,  is proposed in Section \ref{SecAsympTO}. In any case, as we shall see in the next section, to take advantage of the TOs described in this section, a specific signal processing scheme is required in the secondary link.

\subsection{Pre-processing Matrix}\label{SecV2}

In this section, we define the interference alignment condition to be met by the
secondary transmitter and determine a pre-processing matrix satisfying this condition.

\noindent
\begin{definition}[IA condition]\label{DefIACondition}
\emph{Let $\bs{H}_{11} =
\bs{U}_{H_{11}}\bs{\Lambda}_{H_{11}}\bs{V}_{H_{11}}^H$ be an SVD of
$\bs{H}_{11}$ and
\begin{eqnarray}\label{EqR}
\bs{R} & = & \sigma^2_1 \Id_{N_1} + \bs{U}_{H_{11}}^H \bs{H}_{12} \bs{V}_2 \bs{P}_2 \bs{V}_2^H \bs{H}_{12}^H \bs{U}_{H_{11}},
\end{eqnarray}
be the covariance matrix  of the co-channel interference (CCI) plus
noise signal in the primary link. The opportunistic link is said to
satisfy the IA condition if its opportunistic transmission is such
that the primary link achieves the transmission rate of the
equivalent single-user system, which translates mathematically as
}
\begin{equation}\label{EqIACondition} 
\begin{array}{lcl}
\log_{2}\left| \Id_{N_1} + \frac{1}{\sigma^2_1 } \bs{\Lambda}_{H_{11}} \bs{P}_1 \bs{\Lambda}_{H_{11}}^{H}\right| = & & \\ 
\log_{2}\left| \Id_{N_1} +  \bs{R}^{-1} \bs{\Lambda}_{H_{11}} \bs{P}_1 \bs{\Lambda}_{H_{11}}^{H} \right|. & & 
\end{array}
\end{equation}
\end{definition}

\noindent
Our objective is first to find a pre-processing matrix $\bs{V}_2$ that satisfies the
IA condition and then, to tune the PA matrix
$\mathbf{P}_2$ and post-processing matrix $\mb{D}_2$ in order to
maximize the transmission rate for the secondary link.

\noindent
\begin{lemma}[Pre-processing matrix $\bs{V}_2$] \label{LemmaV2}\emph{Let $\bs{H}_{11} = \bs{U}_{H_{11}}\bs{\Lambda}_{H_{11}}\bs{V}_{H_{11}}^H$ be an ordered SVD of
$\bs{H}_{11}$, with  $\bs{U}_{H_{11}}$ and
$\bs{V}_{H_{11}}$, two unitary matrices of size $N_1\times
N_1$ and $M_1 \times M_1$, respectively,  and
$\bs{\Lambda}_{H_{11}}$ an $N_1 \times M_1$ matrix with main
diagonal $\left(\lambda_{H_{11},1},
\ldots,\lambda_{H_{11},\min(N_1,M_1)}\right)$ and zeros on its
off-diagonal, such that $\lambda_{H_{11},1}^2 \geqslant
\lambda_{H_{11},2}^2 \geqslant \ldots \geqslant
\lambda_{H_{11},\min(N_1,M_1)}^2$. Let also the
$N_1 \times M_2$ matrix $\tilde{\bs{H}}
\defn \bs{U}_{H_{11}}^H \bs{H}_{12}$ have a block structure,
\begin{equation}\label{EqBlockHtilde}
\tilde{\bs{H}} =
\begin{array}{rc}
   & \xleftrightarrow{M_2}\\
    \begin{array}{r}
    m_1 \Big\updownarrow \\
    N_1-m_1 \Big\updownarrow
    \end{array}
    &
    \left(\begin{array}{c}
            \tilde{\bs{H}}_1\\
            \tilde{\bs{H}}_2
          \end{array}
          \right)
\end{array}.
\end{equation}
The IA condition (Def. \ref{DefIACondition}) is satisfied independently of the PA matrix $\bs{P}_2$, when the pre-processing matrix $\bs{V}_2$ satisfies the condition:
\begin{equation}\label{EqH1tilde}
                \tilde{\bs{H}}_1 \bs{V}_2 = \bs{0}_{m_1 \times L_2},
\end{equation}}
where $L_2$ is the dimension of the null space of matrix $\tilde{\bs{H}}_1$.
\end{lemma}
\begin{proof} See Appendix \ref{AppProofMatrixR}. \end{proof}
Another solution to the IA condition was given in \cite{IA-Perlaza-08b}, namely $\bs{V}_2 = \bs{H}_{12}^{-1}\bs{U}_{H_{11}}\bar{\bs{P}}_{1}$ for a given diagonal matrix $\bar{\bs{P}}_1 = \diag\left(\bar{p}_{1,1}, \ldots, \bar{p}_{1,M_1}\right)$, with $\bar{p}_{1,n} = \left(\frac{\sigma^2_2}{\lambda_{H_{11}^HH_{11},n}} - \beta\right)^+$, where $\beta$ is the water-level of the primary system (\emph{Th. \ref{ThTelatar95}}) and $n \in \lbrace 1, \ldots, M_1\rbrace$. However, such a solution is more restrictive than (\ref{EqH1tilde}) since it requires $\bs{H}_{12}$ to be invertible and does not hold for the case when $N_i \neq M_j$, $\forall (i,j) \in \lbrace 1,2 \rbrace^2$.

\noindent
Plugging $\bs{V}_2$ from (\ref{EqH1tilde}) into (\ref{EqR}) shows that to guarantee the IA condition (\ref{DefIACondition}), the opportunistic transmitter has to avoid interfering with the $m_1$ dimensions used by the primary transmitter. That is the reason why we refer to our technique as OIA: interference from the secondary user is made orthogonal to the $m_1$ receive dimensions used by the primary link. This is achieved by aligning the interference from the secondary user with the $N_1-m_1$ non-used receive dimensions of the primary link.

\noindent
From Lemma \ref{LemmaV2}, it appears that the $L_2$ columns of matrix
$\bs{V}_{2}$ have to belong to the null space
$\Ker(\tilde{\bs{H}}_1)$ of $\tilde{\bs{H}}_1$ and therefore to the
space spanned by the $\dim  \Ker (\tilde{\bs{H}}_1) =
M_2 - \rank (\tilde{\bs{H}}_1)$ last columns of
matrix $\bs{V}_{\tilde{H}_1}$, where $\tilde{\bs{H}}_1 =
\bs{U}_{\tilde{H}_1} \bs{\Lambda}_{\tilde{H}_1}
\bs{V}_{\tilde{H}_1}^H$ is an SVD of $\bs{\tilde{H}}_{11}$with
$\bs{U}_{\tilde{H}_1}$ and $\bs{V}_{\tilde{H}_1}$ two unitary
matrices of respective sizes $m_1 \times m_1$ and $M_2 \times M_2$, and $\bs{\Lambda}_{\tilde{H}_{1}}$ an $ m_1 \times M_2 $ matrix containing the vector
$(\lambda_{\tilde{H}_{11},1},\ldots,\lambda_{\tilde{H}_{1},\min(m_1, M_2)})$ on its main diagonal and zeros on its off-diagonal, such that $\lambda_{\tilde{H}_{11},1}^2 \geqslant \ldots \geqslant
\lambda_{\tilde{H}_{1},\min(m_1, M_2)}^2$. i.e.,
\begin{equation}\label{EqV21}
\bs{V}_{2} \in \Span \left(\bs{v}_{\tilde{H}_{1}}^{(\rank
(\tilde{\bs{H}}_1)+1)}, \ldots,
\bs{v}_{\tilde{H}_{1}}^{(M_2)}\right).
\end{equation}
Here, for all $i \in  \left\lbrace 1, \ldots, M_2 \right\rbrace$, the column vector $\bs{v}_{\tilde{H}_{1}}^{(i)}$ represents the $i^{th}$ column of matrix $\bs{V}_{\tilde{H}_{1}}$ from the left to the right.

\noindent
In the following, we assume that the $L_2$ columns of the matrix $\bs{V}_{2}$ form an orthonormal basis of the corresponding subspace (\ref{EqV21}), and thus, $\bs{V}_{2}^H\bs{V}_{2} = \Id_{L_2}$.
Moreover, recalling that $\tilde{\bs{H}}_1$ is of size $ m_1 \times M_2$, we would like to point out that:
\begin{itemize}
\item When $m_1 < M_2$,  $\rank(\tilde{\bs{H}}_1)\leq m_1$ and $\dim \Ker(\tilde{\bs{H}}_1) \geq M_2 - m_1$ with equality if and only if $\tilde{\bs{H}}_1$ is full row-rank. This means that there are always at least $M_2 - m_1 >0$ non-null orthogonal vectors in $\Ker(\tilde{\bs{H}}_1)$, and thus, $L_2 = \dim\Ker(\tilde{\bs{H}}_1)$. Consequently, $\bs{V}_{2}$ can always be chosen to be different from the null matrix $\bs{0}_{M_2 \times L_2}$.

\noindent
\item When, $M_2 \leqslant  m_1$, $\rank(\bs{\tilde{H}}_{1})\leq M_2$ and $\dim \Ker(\bs{\tilde{H}}_{1}) \geq 0$,
with equality if and only if $\bs{\tilde{H}}_{1}$ is full
column-rank. This means that there are non-zero vectors in
$\Ker(\bs{\tilde{H}}_{1})$ if and only if $\bs{\tilde{H}}_{1}$ is
not full column-rank. Consequently,  $\bs{V}_{2}$ is a non-zero
matrix if and only if $\bs{\tilde{H}}_{1}$ is not full column-rank,
and again $L_2 = \dim\Ker(\tilde{\bs{H}}_1)$.
\end{itemize}

\noindent
Therefore, the rank of $\bs{V}_{2}$ is given by $L_2=\dim\Ker(\tilde{\bs{H}}_1) \leq M_2$, and it represents the number of transmit dimensions on which the secondary transmitter can allocate power without affecting the performance of the primary user. The following lower bound on $L_2$ holds
\begin{equation}\label{EqL2}
\begin{split}
L_2=\dim\Ker(\tilde{\bs{H}}_1)
&= M_2 - \rank(\tilde{\bs{H}}_1)\\
&\geq M_2 - \min(M_2, m_1)\\
&= \max(0,M_2-m_1)\\
\end{split}
\end{equation}
\noindent
Note that by processing $\bs{s}_2$ with $\bs{V}_{2}$ the resulting
signal $\bs{V}_{2}\bs{s_2}$ becomes orthogonal to the space spanned
by a \emph{subset} of $m_1$  rows of the cross-interference channel matrix
$\tilde{\bs{H}} = \bs{U}_{H_{11}}^H \bs{H}_{12}$. This is the main
difference between the proposed OIA technique and the classical
zero-forcing beamforming (ZFBF) \cite{Paulraj-03}, for which the
transmit signal must be orthogonal to the whole row space of matrix
$\tilde{\bs{H}}$.  In the ZFBF case, the number of transmit dimensions, on which the secondary transmitter can allocate power without affecting the performance of the primary user, is given by $L_{2,BF}=\dim\Ker(\tilde{\bs{H}})= M_2 - \rank(\tilde{\bs{H}})$. Since $\rank(\tilde{\bs{H}}_1) \leq \rank(\tilde{\bs{H}})$, we have $L_{2,BF} \leq L_2$. This inequality, along with the observation that $\Ker(\tilde{\bs{H}}) \subseteq \Ker(\tilde{\bs{H}}_1)$, shows that any opportunity to use a secondary transmit dimension provided by ZFBF is also provided by OIA, thus OIA outperforms ZFBF. In the next section we tackle the problem of
optimizing the post-processing matrix $\bs{D}_2$ to maximize the
achievable transmission rate for the opportunistic transmitter.

\subsection{Post-processing Matrix}\label{SecD2}

Once the pre-processing matrix $\bs{V}_2$ has been adapted to
perform IA according to (\ref{EqV21}), no harmful interference
impairs the primary link. However, the secondary receiver undergoes
the CCI from the primary transmitter. Then, the joint effect of the CCI and noise signals
can be seen as a colored Gaussian noise with covariance matrix
\begin{equation}\label{EqMAICovariance}
\bs{Q} = \bs{H}_{21} \bs{V}_{H_{11}}\bs{P}_1 \bs{V}_{H_{11}}^H \bs{H}_{21}^H  + \sigma^2_2 \Id_{N_2}.
\end{equation}

\noindent
We recall that the
opportunistic receiver has full CSI of all channel matrices, i.e.,
$\bs{H}_{i,j}$, $\forall (i,j) \in \lbrace1,2\rbrace^2$. Given an input covariance matrix $\bs{P}_2$, the mutual information between the input $\bs{s}_2$ and the output $\bs{y}_2=\bs{D}_2\bs{r}_2$ is
\begin{eqnarray}
\nonumber \scriptstyle R_2(\bs{P}_2,\sigma^2_2) & =& \,  \scriptstyle\log_2 \left|\Id_{N_2} + \bs{D}_2\bs{H}_{22} \bs{V}_2 \bs{P}_2 \bs{V}_2^H\bs{H}_{22}^H\bs{D}_2^H \left(\bs{D}_2 \bs{Q} \bs{D}_2^H  \right)^{-1}\right| \\
\label{EqMMI} & \leqslant & \scriptstyle\log_2 \left|\Id_{N_2} + \bs{Q}^{-\frac{1}{2}}\bs{H}_{22} \bs{V}_2 \bs{P}_2 \bs{V}_2^H\bs{H}_{22}^H\bs{Q}^{-\frac{1}{2}} \right|,
\end{eqnarray}
where equality is achieved by a whitening post-processing filter
$\bs{D}_2 = \bs{Q}^{-\frac{1}{2}}$ \cite{Neeser-93}. i.e., the mutual information between the transmitted signal $\bs{s}_2$ and $\bs{r}_2$, is the same as that between $\bs{s}_2$ and $\bs{y}_2 = \bs{D}_2\bs{r}_2$. Note also that expression
(\ref{EqMMI}) is maximized by a zero-mean circularly-symmetric complex Gaussian input $\bs{s}_2$ \cite{Telatar-99}.

\subsection{Power Allocation Matrix Optimization}\label{SecInputCovariance}

In this section, we are interested in finding the input covariance
matrix $\bs{P}_2$ which maximizes the achievable transmission rate
for the opportunistic link, $R_2(\bs{P}_2,\sigma^2_2)$ assuming that
both matrices $\bs{V}_2$ and $\bs{D}_2$ have been set up as
discussed in Sec. \ref{SecV2} and \ref{SecD2}, respectively. More
specifically, the problem of interest in this section is:
\begin{equation}\label{EqOptimizationProblem}
\begin{array}{lc}
\displaystyle\max_{\bs{P}_2} & \scriptstyle\log_2 \left|\Id_{N_2} + \bs{Q}^{-\frac{1}{2}}\bs{H}_{22} \bs{V}_2 \bs{P}_2 \bs{V}_2^H \bs{H}_{22}^H \bs{Q}^{-\frac{1}{2}}\right|\\
\text{s.t.} & \scriptstyle\text{Trace}\left( \bs{V}_2 \bs{P}_2 \bs{V}_2^H \right) \leqslant p_{2,\mathrm{max}}.
\end{array}
\end{equation}
Before solving the optimization problem (OP) in (\ref{EqOptimizationProblem}), we briefly
describe the uniform PA scheme (UPA). The UPA policy can be very
useful not only to relax some information assumptions and decrease
computational complexity at the transmitter but also because it
corresponds to the limit of the optimal PA policy in the high SNR
regime.

\noindent
\subsubsection{Uniform Power Allocation}
In this case, the opportunistic transmitter does not perform any
optimization on its own transmit power. It rather uniformly spreads
its total power among the previously identified TOs. Thus, the PA
matrix $\bs{P}_2$ is assumed to be of the form
\begin{equation}\label{EqPupa}
 \bs{P}_{2,UPA} = \gamma \Id_{L_2},
\end{equation}
where the constant $\gamma$ is  chosen to saturate the transmit
power constraint (\ref{EqPowerConstraints}),
\begin{eqnarray}\label{EqGamma}
     \gamma = \frac{M_2 \: p_{2,\max}}{\tr\left(\bs{V}_2\bs{V}_2^H\right)} = \frac{M_2 p_{2,\max}}{L_2}.
\end{eqnarray}

\noindent
\subsubsection{Optimal Power Allocation}
Here, we tackle the OP formulated in (\ref{EqOptimizationProblem}).
For doing so, we assume that the columns of matrix $\bs{V}_2$ are
unitary and mutually orthogonal. We define the matrix $\bs{K} \defn
\bs{Q}^{-\frac{1}{2}}\bs{H}_{22}\bs{V}_2$, where $\bs{K}$ is an
$N_2 \times L_2$ matrix. Let $\bs{K} = \bs{U}_K
\bs{\Lambda}_K \bs{V}_K^H$ be an SVD of matrix $\bs{K}$, where the
matrices $\bs{U}_K$ and $ \bs{V}_K$ are unitary matrices with
dimensions $ N_2 \times N_2$ and
$L_2 \times L_2$ respectively. The matrix $\bs{\Lambda}_K$
is an $N_2 \times L_2$ matrix with at most
$\min\left( N_2,L_2\right)$ non-zero singular
values on its main diagonal and zeros in its off-diagonal entries. The entries in the diagonal of the
matrix $\bs{\Lambda}_K$ are denoted by $\lambda_{K,1}, \ldots,
\lambda_{K,\min( N_2,L_2)}$. Finally, the
original OP (\ref{EqOptimizationProblem}) can be rewritten as
\begin{equation}\label{EqOptimizationProblem5}
\begin{array}{ll}
\displaystyle\arg\max_{\bs{P}_2}  & \log_2 \scriptstyle\left|\Id_{N_2} + \bs{\Lambda}_K\bs{V}_K^H \bs{P}_2 \bs{V}_K \bs{\Lambda}_K^H\right|\\
\text{s.t.} & \begin{array}{ll} \text{Trace}\scriptstyle\left(\bs{P}_2\right) &  =  \text{Trace}\scriptstyle\left(\bs{V}_K^H \bs{P}_2 \bs{V}_K\right)\\
                                                                 & \leqslant  M_2 \: p_{2,\max}. \end{array} 
\end{array}
\end{equation}

\noindent
Here, we define the square matrices of dimension $L_2$,
\begin{equation}\label{EqP2Tilde}
\tilde{\bs{P}}_2 \defn \bs{V}_K^H \bs{P}_2 \bs{V}_K,
\end{equation}
and $\bs{\Lambda}_{K^H K} \defn
\bs{\Lambda}_{K}^H \bs{\Lambda}_{K} = \diag\left(\lambda_{K^HK,1},
\ldots, \lambda_{K^HK,L_2}\right)$. Using the new variables
$\tilde{\bs{P}}_2$ and $\bs{\Lambda}_{K^H K}$, we can write that
\begin{equation}\label{EqInequality}
\begin{array}{lcl}
 \scriptstyle\left|\Id_{N_2} + \bs{\Lambda}_K\bs{V}_K^H \bs{P}_2 \bs{V}_K\bs{\Lambda}_K^H \right|
  & = & \scriptstyle\left|\Id_{L_2} + \bs{\Lambda}_{K^HK} \tilde{\bs{P}}_2\right| \\
& \leqslant & \displaystyle\prod_{n = 1}^{L_2} \scriptstyle\left(1 + \lambda_{K^HK,n} \, \tilde{p}_{2,n} \right)\\
\end{array}
\end{equation}
where $\tilde{p}_{2,n}$, with $n \in \lbrace 1, \ldots, L_2
\rbrace$ are the entries of the main diagonal of matrix $\tilde{\bs{P}}_2$. Note that
in (\ref{EqInequality}) equality holds if $\tilde{\bs{P}}_2$ is a
diagonal matrix \cite{Marshall-Olkin-1979}. Thus, choosing
$\tilde{\bs{P}}_2$ to be diagonal maximizes the transmission rate.
Hence, the OP simplifies to
\begin{equation}\label{EqOptimizationProblem6}
\begin{array}{lc}
\displaystyle\max_{\tilde{p}_{2,1}\ldots\tilde{p}_{2,L_2}} & \displaystyle\sum_{n = 1}^{L_2}\log_2\left(1 + \lambda_{K^HK,n} \, \tilde{p}_{2,n} \right)\\
\text{s.t.} & \displaystyle\sum_{n = 1}^{L_2} \tilde{p}_{2,n}  \leqslant M_2 p_{2,\max},
\end{array}
\end{equation}
The simplified optimization problem (\ref{EqOptimizationProblem6})
has eventually a water-filling solution of the form
\begin{equation}\label{EqOpportunisticWaterfilling}
\forall n \in \left\lbrace 1,\ldots, L_2 \right\rbrace, \quad \tilde{p}_{2,n} = \left(\beta_2 - \frac{1}{\lambda_{K^HK,n}}\right)^+,
\end{equation}
where, the water-level $\beta_2$ is determined to saturate the power
constraints in the optimization problem
(\ref{EqOptimizationProblem6}). Once the matrix $\tilde{\bs{P}}_2$ (\ref{EqP2Tilde})
has been obtained using water-filling (\ref{EqOpportunisticWaterfilling}), we define the optimal PA matrix
$\bs{P}_{2,OPA}$ by
\begin{eqnarray}\label{EqOpportunisticPA}
\bs{P}_{2,OPA}  &=&  \diag\left(\tilde{p}_{2,i}, \ldots,\tilde{p}_{2,L_2}\right),
\end{eqnarray}
while the left and right hand factors, $\bs{V}_K$ and $\bs{V}_{K}^H$, of matrix $\tilde{\bs{P}}_2$ in (\ref{EqP2Tilde}) are included in the pre-processing matrix:
\begin{equation}
    \bs{V}_{2,OPA} = \bs{V}_{2} \bs{V}_{K}.
\end{equation}

\noindent
In the next section, we study the achievable transmission rates of the opportunistic link.

\section{Asymptotic Performance of the Secondary link}

In this section, the performance of the secondary link is analyzed in the regime of large number of antennas, which is defined as follows:
\noindent
\begin{definition}[Regime of Large Numbers of Antennas]\label{DefRegimeLNA}
\emph{The regime of large numbers of antennas (RLNA) is defined as follows:}
\begin{itemize}
    \item $\forall i \in \lbrace1,2\rbrace$, $N_i \rightarrow + \infty$;
    \item $\forall j \in \lbrace1,2\rbrace$, $M_j \rightarrow + \infty$;
    \item $\forall (i,j) \in \lbrace1,2\rbrace^2$, $\lim\limits_{\substack{M_j \to + \infty \\  N_i \to + \infty}} \frac{M_j}{N_i} = \alpha_{ij} < + \infty$, and $\alpha_{ij} > 0$ is constant.
\end{itemize}
\end{definition}

\subsection{Asymptotic Number of Transmit Opportunities}\label{SecAsympTO}

In Sec. \ref{SecIAStrategy}, two relevant parameters regarding the performance of the opportunistic system can be identified: the number of TOs ($S$) and the number of transmit dimensions to which the secondary user can allocate power without affecting the performance of the primary user ($L_2$). Indeed, $L_2$ is equivalent to the number of idependent symbols the opportunistic system is able to simultaneously transmit. In the following, we analyze both parameters $S$ and $L_2$ in the RLNA by studying the fractions 
\begin{eqnarray}
\label{EqSInfty} S_{\infty}  &\triangleq& \displaystyle\lim\limits_{\substack{N_1 \to +\infty \\ M_1 \to + \infty}} \frac{S}{M_1}
 \mbox{ and,}\\
\label{EqL2Infty} L_{2,\infty}  &\triangleq & \lim\limits_{\substack{N_1 \to +\infty \\ M_2 \to + \infty}} \frac{L_{2}}{M_2}.
\end{eqnarray}
Using (\ref{EqSumOfS}), the fraction $S_{\infty}$ can be re-written as follows
\begin{eqnarray}
\nonumber S_{\infty}  & =          & \displaystyle\lim\limits_{\substack{N_1 \to +\infty \\ M_2 \to + \infty}} \frac{1}{M_1}\left( N_1 -
														 m_1\right)\\
						& =          & \left(\frac{1}{\alpha_{11}} - m_{1,\infty}\right),
\end{eqnarray}
where, 
\begin{equation}
	m_{1,\infty} \triangleq \displaystyle\lim\limits_{\substack{N_1 \to +\infty \\ M_1 \to + \infty}} \frac{m_1}{M_1}.
\end{equation}

\noindent
As a preliminary step toward determining the expressions of
$S_{\infty}$ and $L_{2,\infty}$, we first show how to find the asymptotic water-level
$\beta_{\infty}$ in the RLNA, and the expression of $m_{1,\infty}$. First, recall from the water-filling
solution (\ref{EqWaterfillingPrimary}) and the power constraint
(\ref{EqPowerConstraints}) that
\begin{equation}\label{EqSumOfPowers}
\frac{1}{M_1} \displaystyle\sum_{n = 1}^{M_1} p_{1,n} =\frac{1}{M_1}
\displaystyle\sum_{n = 1}^{M_1} \left(\beta - \frac{\sigma^2_1}{\lambda_{H_{11}^HH_{11},n}} \right)^+.
\end{equation}
Define the real function $q$  by
\begin{equation}
q(\lambda)= \left|
\begin{array}{ll}
  0, & \mbox{if } \lambda =0, \\
  \left(\beta - \frac{\sigma^2_1}{\lambda} \right)^+, & \mbox{if } \lambda
  >0,
\end{array}
\right.
\end{equation}
which is continuous and bounded on $\Rplus$. (\ref{EqSumOfPowers}) can be rewritten as
\begin{equation}\label{EqSumOfPowers2}
\frac{1}{M_1} \displaystyle\sum_{n = 1}^{M_1} q(\lambda_{H_{11}^H H_{11},n})\\
= \int_{-\infty}^{\infty} q(\lambda ) \:\:
f^{(M_1)}_{H_{11}^HH_{11}}(\lambda) \:\: \d\lambda
\end{equation}
where $f^{(M_1)}_{H_{11}^HH_{11}}$ is the probability density
function associated with the empirical eigenvalue distribution
$F^{(M_1)}_{H_{11}^HH_{11}}$ of matrix $\bs{H}_{11}^H \bs{H}_{11}$.
\noindent
In the RLNA, the empirical eigenvalue distribution
$F^{(M_1)}_{H_{11}^HH_{11}}$ converges almost surely to the
deterministic limiting eigenvalue distribution $F_{H_{11}^HH_{11}}$,
known as the Mar\v{c}enko-Pastur law \cite{Marcenko-67} whose
associated density is
\begin{equation}\label{eq:MarcenkoPastur}
    \textstyle f_{H_{11}^HH_{11}}(\lambda) = \textstyle\left(1-\frac{1}{\alpha_{11}}\right)^{+} \delta(\lambda) +
    \frac{\sqrt{\left(\lambda -a\right)^+ \left(b-\lambda\right)^+}}{2\pi\lambda},
\end{equation}
where, $a =\scriptstyle\left(1- \frac{1}{\sqrt{\alpha_{11}}}\right)^2$ and $b =\scriptstyle\left(1+\frac{1}{\sqrt{\alpha_{11}}} \right)^2$.
Note that the Mar\v{c}enko-Pastur law has a bounded real positive
support $\left\{ \{0\} \cup [a,b] \right\}$ and  $q$ is continuous
and bounded on $\Rplus$. Consequently, in the RLNA, we have the
almost sure convergence of (\ref{EqSumOfPowers2}), i.e.,
\begin{eqnarray}
\nonumber \displaystyle\int_{-\infty}^{\infty} \textstyle q(\lambda ) \:\: f^{(M_1)}_{H_{11}^HH_{11}}(\lambda) \:\: \d\lambda
 \stackrel{{a.s.}}{\longrightarrow}
\displaystyle\int_{-\infty}^{\infty} \textstyle q(\lambda) f_{H_{11}^HH_{11}}(\lambda) \d\lambda
\end{eqnarray}
Thus, in the RLNA (Def. \ref{DefRegimeLNA}), the water-level $\beta_{\infty}$ is the unique solution \cite{Chuah-02} to the
equation
\begin{equation}
\int_{\max(\frac{\sigma^2_1}{\beta},a)}^{b} \scriptstyle\left(\beta - \frac{\sigma^2_1}{\lambda}\right) \frac{\sqrt{\left(\lambda -a\right) \left(b-\lambda\right)}}{2\pi  \lambda} \d\lambda - p_{1,\max}  =0,
\end{equation}
and it does not depend on any specific realization of the channel transfer matrix $\mb{H}_{11}$, but only on the maximum power $p_{1,\max}$ and the receiver noise power $\sigma^2_1$.

\noindent
We can now derive $m_{1,\infty}$. From (\ref{eq:PrimaryResDim}), we have
\begin{eqnarray}\label{EqSumOfS2}
\nonumber m_{1,\infty} &=& \lim\limits_{\substack{N_1 \to +\infty \\ M_1 \to + \infty}} \frac{1}{M_1}\displaystyle\sum_{n = 1}^{M_1} \textstyle\mathds{1}_{\left] \frac{\sigma^2_1}{\beta},+\infty\right[}  (\lambda_{H_{11}^HH_{11},n})\\
\nonumber &=& \lim\limits_{\substack{N_1 \to +\infty \\ M_1 \to + \infty}} \int_{-\infty}^{\infty}  \textstyle\mathds{1}_{\left] \frac{\sigma^2_1}{\beta},+\infty\right[}   (\lambda) \:\: f^{(M_1)}_{H_{11}^HH_{11}}(\lambda) \:\: \d\lambda\\
& \stackrel{{a.s.}}{\longrightarrow}& \int_{\max(a,\frac{\sigma^2_1}{\beta_{\infty}})}^{b}
\textstyle\frac{\sqrt{\left(\lambda -a\right) \left(b-\lambda\right)}}{2\pi \lambda} \:\: \d\lambda.
\end{eqnarray}

\noindent
Thus, given the asymptotic number of transmist dimensions used by the primary link per primary transmit antenna $m_{1,\infty}$,
we obtain the asymptotic number of transmit opportunities per primary transmit antenna $S_{\infty}$ by following (\ref{EqSInfty}), i.e.,
\begin{equation}\label{EqSintegral}
    S_{\infty} =
    \frac{1}{\alpha_{11}}  - \int_{\max(a,\frac{\sigma^2_1}{\beta_{\infty}})}^{b}
\textstyle\frac{\sqrt{\left(\lambda -a\right) \left(b-\lambda\right)}}{2\pi \lambda} \:\: \d\lambda.
\end{equation}

\noindent
From (\ref{EqSBounds}), the following bounds on $S_{\infty}$ hold in the RLNA:
\begin{equation}
\left(\frac{1}{\alpha_{11}} - 1\right)^+ \leq S_{\infty} \leq \frac{1}{\alpha_{12}}
\end{equation}

\noindent
Finally, we give the expression of $L_{2,\infty}$. Recall that $L_2=\dim\Ker(\tilde{\bs{H}}_1)
= M_2 - \rank(\tilde{\bs{H}}_1)$. The rank of $\tilde{\bs{H}}_1$ is given by its number of non-zero singular values, or equivalently by the number of non-zero eigenvalues of matrix $\tilde{\bs{H}}_1^H \tilde{\bs{H}}_1$. Let
$\lambda_{\tilde{H}_1^H \tilde{H}_1,1}, \ldots, \lambda_{\tilde{H}_1^H \tilde{H}_1,M_2}$ denote the eigenvalues of matrix $\tilde{\bs{H}}_1^H \tilde{\bs{H}}_1$. We have
\begin{equation}\label{eq:zeta2inf}
\begin{split}
L_{2,\infty} & = 1- \lim_{N_1,M_2 \rightarrow +\infty} \frac{\rank(\tilde{\bs{H}}_1)}{M_2}\\
&= 1 - \lim_{N_1,M_2 \rightarrow +\infty}  \frac{1}{M_2} \sum_{n = 1}^{M_2} \mathds{1}_{\left] 0,+\infty\right[}  (\lambda_{\tilde{H}_1^H \tilde{H}_1,n})\\
&= 1 - \lim_{N_1,M_2 \rightarrow +\infty}  \int_{-\infty}^{+\infty} \mathds{1}_{\left] 0,+\infty\right[}  (\lambda) f_{\tilde{H}_1^H \tilde{H}_1}^{(M_2)}(\lambda) \d\lambda,
\end{split}
\end{equation}
where $f_{\tilde{H}_1^H \tilde{H}_1}^{(M_2)}(\lambda)$ is the probability density function associated with the empirical eigenvalue distribution $F^{(M_2)}_{\tilde{H}_1^H \tilde{H}_1}$. $\tilde{H}_1$ is of size $m_1 \times M_2$, and the ratio $\frac{M_2}{m_1}$ converges in the RLNA to
\begin{equation}
\tilde{\alpha}_{1} \triangleq \lim_{N_1, M_1, M_2 \rightarrow \infty} \frac{M_2}{m_1} = \frac{\alpha_{12}}{\alpha_{11}m_{1,\infty}} < \infty.
\end{equation}
Thus, in the RLNA, the empirical eigenvalue distribution
$F^{(M_2)}_{\tilde{H}_1^H \tilde{H}_1}$ converges almost surely to the Mar\v{c}enko-Pastur law \cite{Marcenko-67} $F_{\tilde{H}_1^H \tilde{H}_1}$ with associated density
\begin{equation}\label{eq:MarcenkoPastur2}
\begin{split}
    f_{\tilde{H}_1^H \tilde{H}_1}(\lambda) &= \left(1-\frac{1}{\tilde{\alpha}_{1}}\right)^{+} \delta(\lambda) +
    \frac{\sqrt{\left(\lambda -c\right)^+ \left(d-\lambda\right)^+}}{2\pi\lambda},\\
    \mbox{where } c &=\left(1- \frac{1}{\sqrt{\tilde{\alpha}_{1}}}\right)^2 \mbox{ and }
     d =\left(1+ \frac{1}{\sqrt{\tilde{\alpha}_{1}}} \right)^2.
\end{split}
\end{equation}
Using (\ref{eq:MarcenkoPastur2}) in (\ref{eq:zeta2inf}) yields
\begin{equation}
\begin{split}
L_{2,\infty} & \stackrel{{a.s.}}{\longrightarrow} 1 - \int_{-\infty}^{+\infty} \mathds{1}_{\left] 0,+\infty\right[}  (\lambda) f_{\tilde{H}_1^H \tilde{H}_1}(\lambda) \d\lambda \\
& = \int_{-\infty}^{+\infty} \mathds{1}_{\{]-\infty,0]\}}  (\lambda) f_{\tilde{H}_1^H \tilde{H}_1}(\lambda) \d\lambda\\
& = \left(1-\frac{1}{\tilde{\alpha}_{1}}\right)^{+}
\end{split}
\end{equation}
Thus, given the asymptotic water-level $\beta_{\infty}$ for the primary link,
the asymptotic number of TOs per transmit antenna is given by the following expression
\begin{eqnarray}\label{eq:zeta2inf2}
    L_{2,\infty} &=& \left(1- \frac{\alpha_{11}}{\alpha_{12}} m_{1,\infty} \right)^{+}\\
\nonumber 			 &=& \textstyle\left(1- \frac{\alpha_{11}}{\alpha_{12}} \displaystyle\int_{\max(a,\frac{\sigma^2_1}{\beta_{\infty}})}^{b}
\textstyle\frac{\sqrt{\left(\lambda -a\right) \left(b-\lambda\right)}}{ 2\pi \lambda} \:\: \d\lambda \right)^{+}.
\end{eqnarray}

\noindent
Note that the number ($S$) of TOs  as well as the number ($L_2$) of independent symbols that the secondary link can simultaneously transmit are basically determined by the number of antennas and the SNR of the primary system. From (\ref{EqSInfty}), it becomes clear that the higher the SNR of the primary link, the lower the number of TOs. Nonetheless, as we shall see in the numerical examples in Sec. \ref{SecSimulations}, for practical values of SNR there exist a non-zero number of TOs the secondary can always exploit.

\subsection{Asymptotic Transmission Rate of the Opportunistic Link}\label{SecAsymptoticAnalaysis}

In this section, we analyze the behavior of the opportunistic rate per antenna
\begin{equation}\label{EqR2}
\scriptstyle\bar{R}_2(\bs{P}_2, \sigma^2_2) \triangleq  \scriptstyle\frac{1}{N_2}
 \log_2 \left|\Id_{N_2} + \bs{Q}^{-1}\bs{H}_{22} \bs{V}_2 \bs{P}_2 \bs{V}_2^H \bs{H}_{22}^H \right|
\end{equation}
in the RLNA. Interestingly, this quantity can be shown to converge
to a limit, the latter being independent of the realization of
$\mathbf{H}_{22}$. In the present work, we
essentially use this limit to conduct a performance analysis of the
system under investigation but it is important to know that it can
be further exploited, for instance, to prove some properties, or simplify optimization problems
\cite{Dumont-06b}. A key transform for analyzing quantities associated with large systems is
the Stieltjes transform, which we define in App. \ref{AppLemmas}. By exploiting the Stieltjes transform and results from random matrix theory for large systems (See App. \ref{AppLemmas}), it is possible to find the limit of (\ref{EqR2}) in the RLNA. The corresponding result is as follows.

\noindent
\begin{proposition}[Asymptotic Transmission Rate]\label{PropPFLD-08} \emph{Define the matrices
\begin{eqnarray}
\bs{M}_1 & \defn & \bs{H}_{21}\bs{V}_{H_{11}}\bs{P}_1\bs{V}_{H_{11}}^H\bs{H}_{21} ^H\\
\bs{M}_2 & \defn & \bs{H}_{22}\bs{V}_2\bs{P}_2\bs{V}_2^H\bs{H}_{22}^H\\
\bs{M}   & \defn & \bs{M}_1 +\bs{M}_2,
\end{eqnarray}
and consider the system model described in Sec. \ref{SecSystemModel}
with a primary link using the configuration $(\bs{V}_1$, $\bs{D}_1$,
$\bs{P}_1)$ described in Sec. \ref{SecIAPrimaryPerformance}, and a
secondary link with the configuration $(\bs{V}_2$, $\bs{D}_2$,
$\bs{P}_2)$ described in Sec. \ref{SecV2}, \ref{SecD2}, with
$\bs{P}_2$ any PA matrix independent from the noise level
$\sigma^2_2$. Then, in the RLNA (Def. \ref{DefRegimeLNA}),
under the assumption that  $\bs{P}_1$ and
$\bs{V}_2\bs{P}_2\bs{V}_2^H$ have limiting eigenvalue
distributions $F_{P_1}$ and  $F_{V_2 P_2 V_2^H}$ with compact
support, the transmission rate per antenna of the
opportunistic link $(\mathrm{Tx}_2$-$\mathrm{Rx}_2)$ converges almost surely to 
\begin{equation}\label{Eq21}
\bar{R}_{2,\infty} =
\frac{1}{\ln2}\displaystyle\int_{\sigma^2_2}^{+\infty} G_{M_1}\left(-z\right) -
G_{M}\left(-z\right) \mathrm{d}z,
\end{equation}
where, $G_M(z)$ and $G_{M_1}(z)$ are the Stieltjes transforms of the limiting eigenvalue distribution of matrices $\bs{M}$ and $\bs{M}_1$, respectively. $G_M(z)$ and $G_{M_1}(z)$ are obtained by solving the fixed point equations (with unique solution when $z \in \mathds{R}^{-}$ \cite{Silverstein-95a}):
\begin{equation}
G_{M_1}(z)  = \frac{-1}{z - g(G_{M_1}(z))}
\end{equation}
and
\begin{equation}
    G_M(z) = \frac{-1}{z-g(G_M(z))- h(G_M(z))},
\end{equation}
respectively, where the functions  $g(u)$ and  $h(u)$ are defined as follows
\begin{eqnarray}
g(u)& \triangleq &\mathds{E}\left[\frac{p_{1}}{1+ \frac{1}{\alpha_{21}}p_{1} u}\right], \label{Eq2}  \\
h(u)& \triangleq & \mathds{E}\left[\frac{p_2}{1+ \frac{1}{\alpha_{22}} p_2 u}\right]. \label{Eq3}
\end{eqnarray}
with the expectations in (\ref{Eq2}) and (\ref{Eq3}) taken on the random variables $p_{1}$ and $p_{2}$ with distribution $F_{P_1}$ and $ F_{V_2 P_2 V_2^H}$, respectively.}
\end{proposition}
\begin{IEEEproof}For the proof, see Appendix \ref{ProofAsymptoticCapaciy}. \end{IEEEproof}
The (non-trivial) result in Prop. \ref{PropPFLD-08} holds for any
power allocation matrix $\bs{P}_2$ independent of $\sigma^2_2$. In
particular, the case of the uniform power allocation policy
perfectly meets this assumption. This also means that it holds for
the optimum PA policy in the high SNR regime. For low and medium
SNRs, the authors have noticed that the matrix $\bs{P}_{2,OPA}$ is
in general not independent of $\sigma^2_2$. This is because $\bs{P}_2$
is obtained from a water-filling procedure. The corresponding
technical problem is not trivial and is therefore left as an extension of the present work.

\section{Numerical Results}\label{SecSimulations}

\subsection{The Number $S$ of Transmit Opportunities }

\begin{figure}[!ht]
\begin{center}
\includegraphics[width=\linewidth]{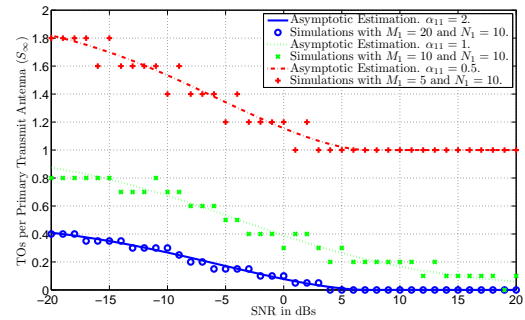}
\caption{\label{FigEstimationS} Fraction of transmit opportunities in the RLNA (Def. \ref{DefRegimeLNA}), i.e., $S_{\infty}$ (\ref{EqSInfty}) as function of the $\mathrm{SNR} = \frac{p_{1,\max}}{\sigma_1^2}$ and  $\alpha_{11} = \frac{M_1}{N_1}$. Simulation results are obtained by using one realization of the matrix $\bs{H}_{11}$ when $N_1 = 10$.}
\end{center}
\end{figure}

As shown in (\ref{EqSInfty}), the number of TOs is a function of the number of antennas and the SNR of the primary link. In Fig. \ref{FigEstimationS}, we plot the number of TOs per transmit antenna $S_{\infty}$ as a function of the SNR for different number of antennas in the receiver and transmitter of the primary link. Interestingly, even though the number of TOs is a non-increasing function of the SNR, Fig. \ref{FigEstimationS} shows that for practical values of the SNR ($10$ - $20$ dBs.) there exists a non-zero number of TOs. Note also that the number of TOs is an increasing function of the ratio ($\alpha_{11} = \frac{M_1}{N_1}$). For instance, in the case $N_1 > M_1$, i.e., $\alpha_{11} > 1$ the secondary transmitters always sees a non-zero number of TOs independently of the SNR of the primary link, and thus, opportunistic communications are always feasible. On the contrary, when $\alpha_{11} \leqslant 1$, the feasibility of opportunistic communications depends on the SNR of the primary link. 

\noindent
Finally, it is important to remark that even though, the analysis of the number of TOs has been done in the RLNA (Def. \ref{DefRegimeLNA}), the model is also valid for finite number of antennas. In Fig. \ref{FigEstimationS}, we have also ploted the number of TOs observed for a given realization of the channel transfer matrix $\bs{H}_{11}$ when $N_1 = 10$ and $\alpha_{11} \in \lbrace \frac{1}{2}, 1,2\rbrace$. Therein, it can be seen how the theretical result from (\ref{EqSInfty}) matches the simulation results. 

\subsection{Comparison between OIA and ZFBF}

We compare our OIA scheme with the zero-forcing beamforming (ZFBF)
scheme \cite{Paulraj-03}. Within this scheme, the pre-processing
matrix $\bs{V}_2$, denoted by $\bs{V}_{2,ZFBF}$, satisfies the
condition
\begin{equation}
    \bs{H}_{12} \bs{V}_{2,ZFBF} = \bs{0}_{N_r,L_2},
\end{equation}
which implies that ZFBF is feasible only in some particular cases
regarding the rank of matrix $\bs{H}_{12}$. For instance, when $M_2
\leqslant N_1$ and $\bs{H}_{12}$ is full column rank, the
pre-processing matrix is the null matrix, i.e., $\bs{V}_{2,ZFBF} =
\bs{0}_{M_2,L_2}$ and thus, no transmission takes place.
On the contrary, in the case of OIA when $M_2 \leqslant N_1$, it is still
possible to opportunistically transmit with a non-null matrix
$\bs{V}_2$ in two cases as shown in Sec. \ref{SecV2}:
\begin{itemize}
\item if $m_1 < M_2$,
\item or if $m_1 \geq M_2$ and $\tilde{\bs{H}}_{1}$ is not full column rank.
\end{itemize}

\noindent
Another remark is that when using ZFBF and both primary and secondary receivers come close, the opportunistic link will observe a significant power reduction since both the targeted and nulling directions become difficult to distinguish. This power reduction will be less significant in the case of OIA since it always holds that $\rank(\bs{V}_2) \geqslant  \rank(\bs{V}_{2,ZFBF})$ thanks to the existence of the additional TOs. Strict equality holds only when $S = \left(\frac{1}{\alpha_{11}} -1\right)^+$. As discussed in Sec. \ref{SecTO}, the number of TOs ($S$) is independent of the position of one receiver with respect to the other. It rather depends on the channel realization $\bs{H}_{11}$ and the SNR of the primary link.

\noindent
In the following, for the ease of presentation, we consider that both primary and secondary devices are equipped with the same number of antennas $N_r = N_1 = N_2$ and $N_t = M_1 = M_2$, respectively. In this scenario, we consider the cases where  $N_t > N_r$ and $N_t \leqslant N_r$.
\noindent
\subsubsection{Case $N_t > N_r$}
\begin{figure}[!ht]
\begin{center}
\includegraphics[width=\linewidth]{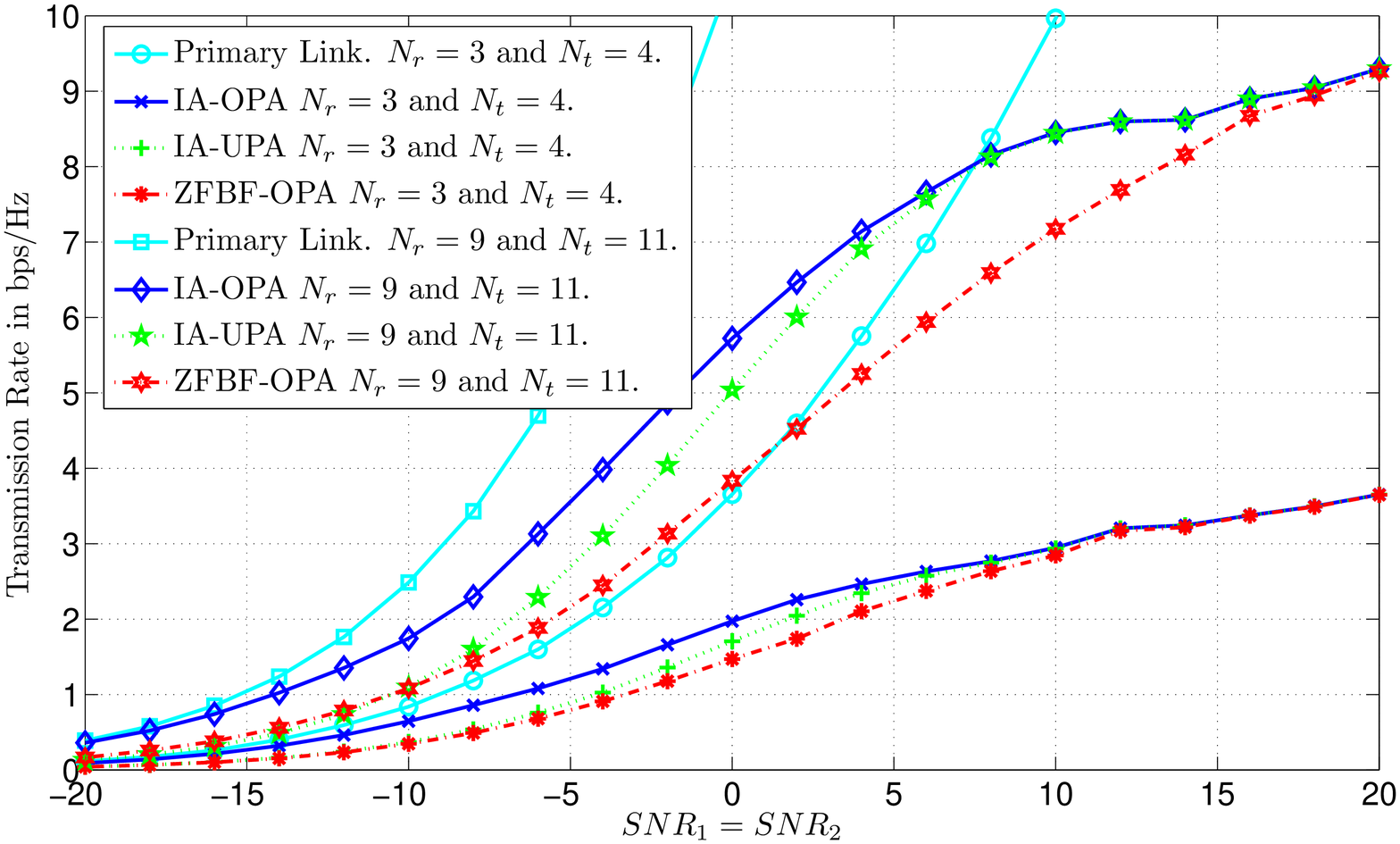}
\caption{\label{FigIA_NtIsNrPlus1} Transmission rate of the opportunistic link obtained by Monte Carlo simulations as a function of the $\mathrm{SNR}_1 = \mathrm{SNR}_2$ when IA and ZFBF are implemented. The number of antennas satisfy $\alpha = \frac{N_t}{N_r} \approx \frac{5}{4}$, with $M_1 = M_2 = N_t$ and  $N_1 = N_2 = N_r \in \left\lbrace 3, 9 \right\rbrace$ and $\mathrm{SNR}_i = \frac{p_{i,\max}}{\sigma^2_1}$, for all $i \in \lbrace 1,2\rbrace$.}
\end{center}
\end{figure}

In Fig. \ref{FigIA_NtIsNrPlus1}, we consider the case where $\alpha
\approx \frac{5}{4}$, with $N_r \in \left\lbrace 3, 9 \right\rbrace$. In
this case, we observe that even for a small number of antennas, the
OIA technique is superior to the classical ZFBF. Moreover, the
higher the number of antennas, the higher the difference
between the performance of  both techniques. An important remark
here is that, at high SNR, the performance of ZFBF and OIA
is almost identical. This is basically because at high SNR, the number of TOs tends to its lower bound $N_t - N_r$ (from (\ref{EqSBounds})), which coincides with the number of spatial directions to which ZFBF can avoid intefering. Another remark is that both UPA and OPA schemes perform identically at high SNR.

\noindent
\subsubsection{Case $N_t \leqslant N_r$}
\begin{figure}[!ht]
\begin{center}
\includegraphics[width=\linewidth]{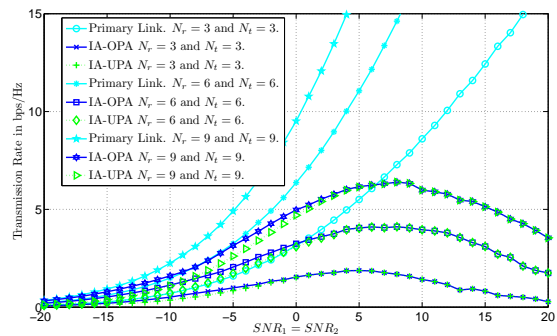}
\caption{\label{FigIA_NtEqualNr} Transmission rate of the opportunistic link obtained by Monte Carlo simulations as a function of the $\mathrm{SNR}_1 = \mathrm{SNR}_2$. The number of antennas satisfy $M_1 = M_2 = N_t$ and  $N_1 = N_2 = N_r$, with $N_t = N_r$, and $N_r \in \left\lbrace 3, 6, 9 \right\rbrace$ and  $\mathrm{SNR}_i = \frac{p_{i,\max}}{\sigma_i^2}$, for all $i \in \lbrace 1,2\rbrace$.}
\end{center}
\end{figure}

\begin{figure}[!ht]
\begin{center}
\includegraphics[width=\linewidth]{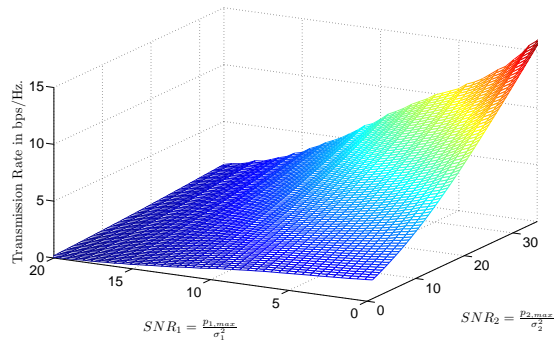}
\caption{\label{FigOIA3D} Transmission rate of the opportunistic link obtained by Monte Carlo simulations as a function of the  $\mathrm{SNR}_i = \frac{p_{i,\max}}{\sigma_i^2}$, with $i \in \lbrace 1, 2 \rbrace$. The number of antennas satisfy $M_1 = M_2 = N_t$ and $N_1 = N_2 = N_r$, with $N_r = N_t = 4$.}
\end{center}
\end{figure}

In this case, the ZFBF solution is not feasible and thus, we focus only on the OIA solution. In Fig. \ref{FigIA_NtEqualNr}, we plot the transmission rate for the case where $N_r = N_t \in
\left\lbrace 3, 6, 9\right\rbrace$. We observe that at high SNR for
the primary link and small number of antennas, the uniform PA
performs similarly as the optimal PA. For a higher number of
antennas and low SNR in the primary link, the difference between
the uniform and optimal PA is significant. To show the impact of the SINR of both primary and secondary links on the opportunistic transmission rate, we present Fig.\ref{FigOIA3D}. Therein, it can be seen clearly that the transmission rate in the opportunistic link is inversely proportional to the SNR level at the primary link. This is due to the lack of TOs as stated in Sec. \ref{SecTO}. For the case when $N_r < N_t$ with strict inequality, an opportunistic transmission takes place only if $N_r - N_t \leqslant S$ and $\bs{\tilde{H}}_{11}$ is not full column rank. Here, the behaviour of the opportunistic transmission rate is similar to the case $N_r = N_t$ with the particularity that the opportunistic transmission rate reaches zero at a lower SNR level. As in the previous case, this is also a consequence of the number of available TOs.
 
\subsection{Asymptotic Transmission Rate}

\begin{figure}[!ht]
\begin{center}
\includegraphics[width=\linewidth]{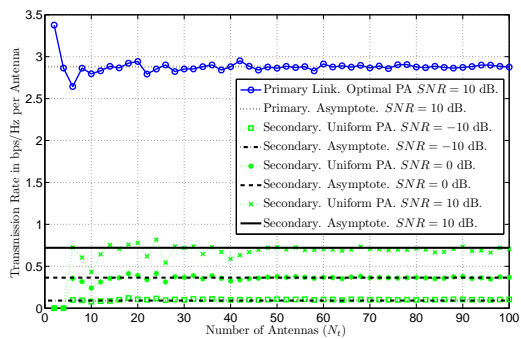}
\caption{\label{FigAsymptoticCapacity} Asymptotic transmission rates per antenna of the opportunistic link  as a function of the number of antennas when $N_r = N_t$ using uniform PA at different SNR levels $\mathrm{SNR}_i = \frac{p_{i,\max}}{\sigma^2_i}$. Simulation results are obtained using one channel realization  for matrices $\bs{H}_{ij}$ $\forall (i,j) \in \lbrace 1, 2 \rbrace^2$ and theoretical results using Prop. \ref{PropPFLD-08}, }
\end{center}
\end{figure}

In Fig. \ref{FigAsymptoticCapacity}, we plot both primary and
secondary transmission rates for a given realization of  matrices
$\bs{H}_{i,j}$ $\forall (i,j) \in \lbrace 1,2 \rbrace^2$. We also
plot the asymptotes obtained from Prop. \ref{PropPFLD-08}
considering UPA in the secondary link and the optimal PA of the
primary link (\ref{EqWaterfillingPrimary}). We observe that in both
cases the transmission rate converges rapidly to the asymptotes even
for a small number of antennas. This shows that Prop.
\ref{PropPFLD-08} constitutes a good estimation of the achievable
transmission rate for the secondary link even for finite number of
antennas. We use Prop. \ref{PropPFLD-08} to compare the asymptotic
transmission rate of the secondary and primary link. The asymptotic
transmission rate of the primary receiver corresponds to the
capacity of a single user $N_t \times N_r$ MIMO link whose asymptotes
are provided in \cite{Foschini-96}. From Fig.
\ref{FigAsymptoticCapacity}, it becomes evident how the secondary
link is able to achieve transmission rates of the same order as the
primary link depending on both its own SNR and that of the primary
link.

\section{Conclusions}\label{SecConclusions}

In this paper, we proposed a technique to recycle spatial directions
left unused by a primary MIMO link, so that they can be re-used by
secondary links. Interestingly, the number of spatial directions can
be evaluated analytically and shown to be sufficiently high to allow
a secondary system to achieve a significant transmission rate. We
provided a signal construction technique to exploit those spatial
resources and a power allocation policy which maximizes the
opportunistic transmission rate. Based on our asymptotical analysis,
we show that this technique allows a secondary link to achieve
transmission rates of the same order as those of the primary link,
depending on their respective SNRs. To mention few interesting
extensions of this work, we recall that our solution concerns only
two MIMO links. The case where there exists several opportunistic
devices and/or several primary devices remains to be studied in
details. More importantly, some information assumptions could
be relaxed to make the proposed approach more practical. This remark
concerns CSI assumptions but also behavioral assumptions. Indeed, it
was assumed that the precoding scheme used by the primary
transmitter is capacity-achieving, which allows the secondary
transmitter to predict how the secondary transmitter is going to
exploit its spatial resources. This behavioral assumption
could be relaxed but some spatial sensing mechanisms should be
designed to know which spatial modes are effectively used by the
secondary transmitter, which could be an interesting extension of
the proposed scheme.

\appendices

\section{Proof of Lemma \ref{LemmaV2}}\label{AppProofMatrixR}

Here, we prove \emph{Lemma \ref{LemmaV2}} which states that:  if a matrix
$\bs{V}_2$ satisfies the condition $ \tilde{\bs{H}}_1 \bs{V}_2 =
\bs{0}_{(N_1-S) \times L_2}$ then it meets the IA condition
(\ref{DefIACondition}).

\noindent
\begin{IEEEproof}
Let $\bs{H}_{11} = \bs{U}_{H_{11}} \bs{\Lambda}_{H_{11}}\bs{V}^H_{H_{11}}$ be a sorted SVD of matrix $\bs{H_{11}}$, with $\bs{U}_{H_{11}}$ and $\bs{V}_{H_{11}}$, two unitary matrices of sizes $N_1\times N_1$ and $M_1 \times M_1$, respectively, and $\bs{\Lambda}_{H_{11}}$ an $N_1 \times M_1$ matrix with main diagonal $\left(\lambda_{H_{11},1}, \ldots,
\lambda_{H_{11},\min(N_1,M_1)}\right)$ and zeros on its off-diagonal, such that  $\lambda_{H_{11},1}^2 \geqslant
\lambda_{H_{11},2}^2 \geqslant \ldots \geqslant
\lambda_{H_{11},\min(N_1,M_1)}^2$. Given that the singular values of the matrix $\bs{H}_{11}$ are sorted, we can write matrix $\bs{\Lambda}_{H_{11}} \bs{P}_1 \bs{\Lambda}_{H_{11}}^{H}$ as a block matrix,
\begin{equation}\label{EqLPLBlock}
    \scriptstyle\bs{\Lambda}_{H_{11}} \bs{P}_1 \bs{\Lambda}_{H_{11}}^{H}
     = \left(\begin{array}{cc} \scriptstyle\bs{\Psi} & \scriptstyle\bs{0}_{m_1 \times (N_1 - m_1) } \\
     \scriptstyle\bs{0}_{(N_1 - m_1) \times m_1 } & \scriptstyle\bs{0}_{(N_1 - m_1) \times (N_1 - m_1)}\end{array} \right),
\end{equation}
where the diagonal matrix $\bs{\Psi}$ of size $m_1 \times m_1$ is
$\bs{\Psi} = \diag\left(\lambda_{H_{11},1}^2\:p_{1,1},
\ldots, \lambda_{H_{11},m_1}^2\:p_{1,m_1}\right)$.

\noindent
Now let us split the interference-plus-noise covariance matrix
(\ref{EqR}) as:
\begin{equation}\label{EqInvRBlock}
    \scriptstyle\bs{R} = \begin{array}{lcc}
    & \xlongleftrightarrow{m_1} & \xlongleftrightarrow{\scriptstyle N_1-m_1}\\
    \begin{array}{r}
    \scriptstyle m_1 \Big\updownarrow \\
    \scriptstyle N_1-m_1 \Big\updownarrow \\
  \end{array}
  &
  \left(\begin{array}{l}
    \scriptstyle \bs{R}_{1} + \sigma_1^2 \Id_{m_1} \\
    \scriptstyle \bs{R}_{2}^H
  \end{array}\right.
  &
    \left.\begin{array}{l}
    \scriptstyle\bs{R}_{2}\\
    \scriptstyle\bs{R}_{3} + \sigma_1^2 \Id_{N_1-m_1}
  \end{array}\right),
\end{array}
\end{equation}
where $\left( \bs{R}_1 + \sigma^2_1 \Id_{m_1}\right)$ and $\left(\bs{R}_3 +
\sigma^2_1\Id_{N_1-m_1}\right)$ are invertible Hermitian matrices, and
matrices $\bs{R}_1$, $\bs{R}_2$ and $\bs{R}_3$ are defined from
(\ref{EqR}) and (\ref{EqBlockHtilde}) as
\begin{eqnarray}
\bs{R}_1 &\triangleq&  \tilde{\bs{H}}_{1} \bs{V}_2 \bs{P}_2 \bs{V}_2^H \tilde{\bs{H}}_{1}^H, \label{EqRCond1} \\
\bs{R}_2 &\triangleq&  \tilde{\bs{H}}_{1} \bs{V}_2 \bs{P}_2 \bs{V}_2^H \tilde{\bs{H}}_{2}^H, \label{EqRCond2} \\
\bs{R}_3 &\triangleq&  \tilde{\bs{H}}_{2} \bs{V}_2 \bs{P}_2 \bs{V}_2^H \tilde{\bs{H}}_{2}^H. \label{EqRCond3}
\end{eqnarray}
Now, by plugging expressions (\ref{EqLPLBlock}) and (\ref{EqInvRBlock})  in (\ref{EqIACondition}), the IA condition can be rewritten as follows:
\begin{equation}\label{EqRCond4}
\begin{array}{l}
\scriptstyle\log_2\left|\sigma^2_{1} \Id_{m_1} +  \bs{\Psi} \right| - \log_2\left| \sigma^2_{1} \Id_{N_1}\right| = \log_2\left|\bs{R}_1 + \sigma_1^2 \Id_{m_1} + \bs{\Psi} \right| \\
\scriptstyle - \log_2\left| \bs{R}_1 + \sigma_1^2 \Id_{m_1} \right|  -\\
 \scriptstyle \log_2\left(\frac{\left| \bs{R}_3+ \sigma_1^2 \Id_{N_1-m_1} - \bs{R}_2^H \left(\bs{R}_1 + \sigma_1^2 \Id_{m_1} \right)^{-1}\bs{R}_2 \right|}{\left| \bs{R}_3 + \sigma_1^2 \Id_{N_1-m_1} - \bs{R}_2^H \left(\bs{R}_1+ \sigma_1^2 \Id_{m_1} + \bs{\Psi}\right)^{-1}\bs{R}_2 \right|}\right).
\end{array}
\end{equation}
Note that there exists several choices for the submatrices $\bs{R}_1$, $\bs{R}_2$, and $\bs{R}_3$ allowing the equality in (\ref{EqRCond4}) to be met. We see that a possible choice in order to meet the IA condition is $\bs{R}_1 = \bs{0}$, $\bs{R}_2 = \bs{0}$, independently of the matrix $\bs{R}_3$. Thus, from (\ref{EqRCond1}) and (\ref{EqRCond2}) we have $\bs{R}_1 = \bs{0}$ and $\bs{R}_2 = \bs{0}$ by imposing the condition $\tilde{\bs{H}}_{1} \bs{V}_2 = \bs{0}_{m_1 \times L_2}$, for any given PA matrix $\bs{P}_2$, which concludes the proof.
\end{IEEEproof}

\section{Definitions}\label{AppLemmas}
In this appendix, we present useful definitions and previous results used in the proofs of Appendix \ref{ProofAsymptoticCapaciy}.

\noindent
\begin{definition}\label{DefTransforms} \emph{Let $\bs{X}$ be an $n\times n$ random matrix with empirical eigenvalue
distribution function $F^{(n)}_X$. We define the following transforms associated with the distribution $F^{(n)}_X$, for $z\in \mathds{C}^+=\{ z \in \mathds{C}: \im (z)>0\}$:
\begin{eqnarray}
\label{Eq19} \mbox{Stieltjes transform:}\, \scriptstyle G_X\left(z\right) &\defn& \int_{-\infty}^{\infty} \textstyle\frac{1}{t-z} \scriptstyle\mathrm{d}F^{(n)}_X(t),\\
\scriptstyle \Upsilon_{X}(z) &\defn& \int_{-\infty}^{\infty} \textstyle\frac{z t}{1- zt}\scriptstyle\mathrm{d}F^{(n)}_{X}(t), \label{Eq53}\\
\mbox{S-transform:}\,   \scriptstyle S_{X}(z) &\defn& \textstyle \frac{1+z}{z}\scriptstyle\Upsilon_{X}^{-1}(z), \label{Eq52}
\end{eqnarray}
where the function $\Upsilon_{X}^{-1}(z)$ is the reciprocal function of $\Upsilon_{X}(z)$, i.e.,
\begin{equation}\label{EqUpsilonReciprocal}
\Upsilon_{X}^{-1}(\Upsilon_{X}(z))= \Upsilon_{X}(\Upsilon_{X}^{-1}(z)) = z.
\end{equation}}
\end{definition}
From (\ref{Eq19}) and (\ref{Eq53}), we obtain the following
relationship between the function $\Upsilon_X(z)$ (named
$\Upsilon$-transform in \cite{Verdu-Book-04}) and the Stieltjes
transform $G_X(z)$,
\begin{equation}\label{Eq53a}
\Upsilon_{X}(z) = -1 - \frac{1}{z}G_{X}\left(\frac{1}{z}\right).
\end{equation}

\section{Proof of Proposition \ref{PropPFLD-08}}\label{ProofAsymptoticCapaciy}

In this appendix, we provide a proof of Prop. \ref{PropPFLD-08} on the asymptotic expression of the opportunistic transmission rate per antenna, defined by $$ \bar{R}_{2,\infty}(\bs{P}_2, \sigma^2) \triangleq
\displaystyle\lim\limits_{\substack{\forall (i,j) \in \lbrace1,2\rbrace^2, \, N_i, M_j \rightarrow \infty \\ \forall (i,j) \in \lbrace1,2\rbrace^2, \,\frac{M_j}{N_i} \rightarrow \alpha_{ij} < \infty}} \bar{R}_{2}(\bs{P}_2, \sigma^2).$$
First, we list the steps of the proof and then we present a detailed development for each of them:
\begin{enumerate}
    \item Step 1: Express $\frac{\partial \bar{R}_{2,\infty}(\bs{P}_2,\sigma^2_2)}{\partial\sigma^2_2}$ as function of the Stieltjes transforms $G_{M_1}(z)$ and $G_{M}(z)$,
    \item Step 2: Obtain $G_{M_1}(z)$,
    \item Step 3: Obtain $G_{M}(z)$,
    \item Step 4: Integrate $\frac{\partial \bar{R}_{2,\infty}(\bs{P}_2,\sigma^2_2)}{\partial\sigma^2_2}$ to obtain  $\bar{R}_{2,\infty}(\bs{P}_2,\sigma^2_2)$.
\end{enumerate}

\noindent
\textbf{Step 1: Express $\frac{\partial \bar{R}_{2,\infty}(\bs{P}_2,\sigma^2_2)}{\partial\sigma^2_2}$ as a function of the Stieltjes transforms $G_{M_1}(z)$ and $G_{M}(z)$.}

\noindent
Using (\ref{EqMMI}) and  (\ref{EqMAICovariance}), the opportunistic
rate per receive antenna  $\bar{R}_2$ can be re-written as follows
\begin{eqnarray}
\label{Eq13} \scriptstyle \bar{R}_2(\bs{P}_2, \sigma^2_2) &=& \scriptstyle \frac{1}{N_2} \log_2 \left|\Id_{N_2} + \bs{Q}^{-\frac{1}{2}}\bs{H}_{22} \bs{V}_2 \bs{P}_2 \bs{V}_2^H \bs{H}_{22}^H \bs{Q}^{-\frac{1}{2}} \right|\\
\nonumber & = & \scriptstyle \frac{1}{N_2} \log_2 \left|\sigma^2_2 \Id_{N_2} + \bs{M}_1 + \bs{M}_2\right|-  \scriptstyle \frac{1}{N_2} \log_2 \left|\sigma^2_2 \Id_{N_2} + \bs{M}_1\right|,
\end{eqnarray}
with $\bs{M}_{1} \defn \bs{H}_{21} \bs{V}_{H_{11}}\bs{P}_1 \bs{V}_{H_{11}}^H \bs{H}_{21}^H$, $\bs{M}_{2}  \defn \bs{H}_{22}\bs{V}_2 \bs{P}_2\bs{V}_2^{H} \bs{H}_{22}^H$, and $\bs{M} = \bs{M}_1 + \bs{M}_2$. Matrices $\bs{M}$ and $\bs{M}_1$ are Hermitian Gramian matrices with eigenvalue decomposition $\bs{M} = \bs{U}_M \bs{\Lambda}_M
\bs{U}_M^H$ and $\bs{M}_1 = \bs{U}_{M_1} \bs{\Lambda}_{M_1}
\bs{U}_{M_1}^H$, respectively. Matrix $\bs{U}_M$ and $\bs{U}_{M_1}$
are $N_2 \times N_2$ unitary matrices, and $\bs{\Lambda}_M = \diag
(\lambda_{M,1}, \ldots,  \lambda_{M,N_2})$ and
$\bs{\Lambda}_{M_1}=\diag(\lambda_{M_1,1}, \ldots,  \lambda_{M_1,N_2})$
are square diagonal matrices containing the eigenvalues of the
matrices $\bs{M}$ and $\bs{M}_{1}$ in decreasing order. Expression
(\ref{Eq13}) can be written as
\begin{eqnarray}\label{Eq14}
\scriptstyle \bar{R}_2(\bs{P}_2, \sigma^2_2)
& = & \scriptstyle \frac{1}{N_2} \displaystyle\sum_{i=1}^{N_2} \scriptstyle \log_2\left(\sigma^2_2 + \lambda_{M,i} \right) - 
\scriptstyle \log_2\left(\sigma^2_2 + \lambda_{M_1,i} \right)\\
\nonumber &= & \displaystyle\int \scriptstyle\log_2 \left(\lambda+ \sigma_2^2\right)\mathrm{d}F_{M}^{(N_2)}(\lambda) -
\log_2 \left(\lambda + \sigma_2^2\right)\mathrm{d}F_{M_1}^{(N_2)}(\lambda) \\
\nonumber  & \stackrel{a.s}{\rightarrow}&   \displaystyle\int  \scriptstyle \log_2 \left(\lambda+ \sigma^2_2\right)\mathrm{d}F_{M}(\lambda)
- \displaystyle\int  \scriptstyle \log_2 \left(\lambda + \sigma^2_2\right)\mathrm{d}F_{M_1}(\lambda),
\end{eqnarray}
where $F_{M}^{(N_2)}$ and $F_{M_1}^{(N_2)}$ are respectively the
empirical eigenvalue distributions of matrices $\bs{M}$ and
$\bs{M}_1$ of size $N_2$, that converge almost surely to the
asymptotic eigenvalue distributions $F_{M}$ and $F_{M_1}$,
respectively. $F_{M}$ and $F_{M_1}$ have a compact support. Indeed
the empirical eigenvalue distribution of Wishart matrices
$\bs{H}_{ij}\bs{H}_{ij}^H$ converges almost surely to the compactly
supported Mar\v{c}enko-Pastur law, and by assumption, matrices
$\bs{V}_i \bs{P}_i \bs{V}_i^H$, $i \in \lbrace1, 2\rbrace$ have a limit eigenvalue
distribution with a compact support. Then by \emph{Lemma~5} in
\cite{Fawaz-08}, the asymptotic eigenvalue distribution of
$\bs{M}_1$ and $\bs{M}_2$ have a compact support. The logarithm
function being continuous, it is bounded on the compact supports of
the asymptotic eigenvalue distributions of $\bs{M}_1$ and $\bs{M}$,
therefore, the almost sure convergence in (\ref{Eq14}) could be
obtained by using the bounded convergence theorem
\cite{Bartle-1995}.

\noindent
From (\ref{Eq14}), the derivative of the asymptotic rate
$\bar{R}_{2,\infty}(\bs{P}_2, \sigma^2)$ with respect to the noise
power $\sigma^2_2$ can be written as
\begin{eqnarray}\label{Eq20}
\scriptstyle\frac{\partial}{\partial\sigma^2_2} \bar{R}_{2,\infty}(\bs{P}_2, \sigma^2_2)
&=& \scriptstyle\frac{1}{\ln2} \left(\displaystyle\int \scriptstyle \frac{1}{\sigma^2_2 + \lambda} \mathrm{d}F_{M}(\lambda) - \displaystyle\int \scriptstyle\frac{1}{\sigma^2_2 + \lambda} \mathrm{d}F_{M_1}(\lambda) \right) \nonumber\\
&=& \scriptstyle \frac{1}{\ln2} \scriptstyle \left( G_M\left(-\sigma^2_2\right) - G_{M_1}\left(-\sigma^2_2\right) \right).
\end{eqnarray}
where $G_{M}\left(z\right)$  and $G_{M_1}\left(z\right)$ are the Stieltjes transforms  of the asymptotic eigenvalue distributions $F_{M}$ and $F_{M_1}$, respectively.

\noindent
\textbf{Step 2: Obtain $G_{M_1}(z)$}

\noindent
Matrix $\bs{M}_1$ can be written as
\begin{equation}
\bs{M}_1   = \sqrt{\alpha_{21}} \bs{H}_{21} \bs{V}_{H_{11}} \frac{\bs{P}_1}{\alpha_{21}} \bs{V}_{H_{11}}^H \bs{H}_{21}^H \sqrt{\alpha_{21}}.
\end{equation}
The entries of the $N_2 \times M_1$ matrix $\sqrt{\alpha_{21}}
\bs{H}_{21}$ are zero-mean i.i.d. complex Gaussian with variance
$\frac{\alpha_{21}}{M_1}=\frac{1}{N_2}$, thus $\sqrt{\alpha_{21}} \bs{H}_{21}$
is  bi-unitarily invariant. Matrix $\bs{V}_{H_{11}}$ is unitary,
consequently $\sqrt{\alpha_{21}} \bs{H}_{21}\bs{V}_{H_{11}}$ has the same
distribution as $\sqrt{\alpha_{21}} \bs{H}_{21}$, in particular its
entries are i.i.d. with mean zero and variance $\frac{1}{N_2}$. From
(\ref{EqWaterfillingPrimary}), $\frac{\bs{P}_1}{\alpha_{21}}$ is
diagonal, and by assumption it has a limit eigenvalue distribution
$F_{\frac{\bs{P}_1}{\alpha_{21}}}$. Thus we can apply \emph{Theorem~1.1} in
\cite{Silverstein-95a} to $\bs{M}_1$, in the particular case where
$\bs{A} = \bs{0}_{N_2}$ to obtain the Stieltjes transform of the
asymptotic eigenvalue distribution of matrix $\bs{M}_1$
\begin{eqnarray}
\nonumber \scriptstyle G_{M_1} \left(z \right)  &= & \scriptstyle G_{\bs{0}_{N_2}} \left(z- \alpha_{21}  \displaystyle\int \scriptstyle\frac{\lambda}{1+\lambda G_{M_1}\left(z \right)} \d F_{\frac{P_1}{\alpha_{21}}} (\lambda) \right)\\
\nonumber &=& \scriptstyle G_{\bs{0}_{N_2}} \left(z- \alpha_{21} \displaystyle\int_{-\infty}^{\infty}  \scriptstyle\frac{\lambda}{1+\lambda G_{M_1}\left(z \right)} \alpha_{21} f_{P_1} (\alpha_{21} \lambda) \d\lambda \right)\\
\nonumber &=& \scriptstyle G_{\bs{0}_{N_2}} \left(z- \displaystyle\int_{-\infty}^{\infty}  \scriptstyle\frac{t}{1+\frac{t}{\alpha_{21}} G_{M_1}\left(z \right)} f_{P_1} (t) \mathrm{d}t \right)\\
\label{Eq39} &=& \scriptstyle G_{\bs{0}_{N_2}} \left(\: z -  g( G_{M_1}(z) ) \: \right),
\end{eqnarray}
where the function $g(u)$ is defined by
\begin{eqnarray}
\nonumber g(u) & \triangleq & \displaystyle\int_{-\infty}^{\infty} \textstyle \frac{t}{1+\frac{t}{\alpha_{21}} u} f_{P_1} (t) \mathrm{d}t = \mathds{E}\textstyle \left[\frac{t}{1+ \frac{1}{\alpha_{21}}t u}\right],
	\end{eqnarray}
where the random variable $t$ follows the c.d.f. $F_{P_1}$.

\noindent
The square null matrix $\bs{0}$ has an asymptotic eigenvalue distribution $F_{\bs{0}}(\lambda) = \mu(\lambda)$. Thus, its Stieltjes transform is
\begin{equation}\label{Eq40}
G_{\bs{0}} (z)= \displaystyle\int_{-\infty}^{\infty} \frac{1}{\lambda-z}\delta(\lambda) \mathrm{d}\lambda = -\frac{1}{z}.
\end{equation}
Then, using expressions (\ref{Eq39}) and (\ref{Eq40}), we obtain
\begin{equation}\label{Eq47}
\begin{split}
G_{M_1}(z)  &= \frac{-1}{z - g(G_{M_1}(z))}.
\end{split}
\end{equation}

\noindent
Expression (\ref{Eq47}) is a fixed-point equation with unique solution when $z \in \mathds{R}^{-}$ \cite{Silverstein-95a}.

\noindent
\textbf{Step 3: Obtain $G_{M}(z)$}
Recall that
\begin{equation}
\textstyle\bs{M} \triangleq \textstyle\bs{H}_{22}\bs{V}_2 \bs{P}_2\bs{V}_2^{H} \bs{H}_{22}^H + \bs{H}_{21} \bs{V}_{H_{11}}\bs{P}_1 \bs{V}_{H_{11}}^H \bs{H}_{21}^H
\end{equation}
To obtain the Stieltjes transform $G_M$, we apply \emph{Theorem~1.1} in  \cite{Silverstein-95a} as in Step 2:
\begin{equation}\label{EqG_M}
G_M\left(z\right)= G_{M_2}\left( \: z- g(G_{M}(z))\: \right).
\end{equation}

\noindent
To obtain the Stieltjes transform $G_{M_2}$ of the asymptotic eigenvalue distribution function of the matrix $\bs{M}_2 = \bs{H}_{22} \bs{V}_2 \bs{P}_2 \bs{V}_2^{H} \bs{H}_{22}^H$, we first express its $S$-transform as
\begin{eqnarray}
\nonumber S_{M_2}(z) & = & S_{H_{22} V_{2} P_2 V_{2}^H H_{22}^H}(z) \\
\nonumber & = & S_{\sqrt{\alpha_{22}}H_{22} V_{2} \frac{P_2}{\alpha_{22}} V_{2}^H H_{22}^H\sqrt{\alpha_{22}}}(z),
\end{eqnarray}
and by \emph{Lemma~1 in \cite{Fawaz-08}}:
\begin{eqnarray}
\nonumber \scriptstyle S_{M_2}(z) & = & \scriptstyle\left(\frac{z + 1}{z + \alpha_{22}} \right)  S_{\sqrt{\alpha_{22}}H_{22}^H H_{22}\sqrt{\alpha_{22}} V_{2} \frac{P_2}{\alpha_{22}} V_{2}^H }(\frac{z}{\alpha_{22}}),
\end{eqnarray}
and by \emph{Theorem~1 in \cite{Mueller-2002}}:
\begin{eqnarray}
\nonumber \scriptstyle S_{M_2}(z) & = & \scriptstyle\left(\frac{z + 1}{z + \alpha_{22}} \right)  S_{\sqrt{\alpha_{22}}H_{22}^H H_{22}\sqrt{\alpha_{22}}}\left(\frac{z}{\alpha_{22}}\right) S_{V_{2} \frac{\bs{P}_2}{\alpha_{22}} V_{2}^H}\left(\frac{z}{\alpha_{22}}\right) \\
\nonumber & = & \scriptstyle\left(\frac{z + 1}{z + \alpha_{22}} \right) \left(\frac{1}{ 1 + \alpha_{22}\frac{z}{\alpha_{22}}} \right) S_{V_{2} \frac{P_2}{\alpha_{22}} V_{2}^H}\left(\frac{z}{\alpha_{22}}\right)\\
\label{EqG_M2} & = &  \scriptstyle\left(\frac{1}{z+\alpha_{22}} \right)  S_{V_{2} \frac{\bs{P}_2}{\alpha_{22} }V_{2}^H}\left(\frac{z}{\alpha_{22}}\right)
\end{eqnarray}

\noindent
The $S$-transforms $ S_{M_2}(z)$ and $ S_{V_{2} \bs{P}_2
V_{2}^H}\left(\frac{z}{\alpha}\right)$ in expression (\ref{EqG_M2})
can be written as functions of their $\Upsilon$-transforms:
\begin{eqnarray}
\label{EqS_M2} \scriptstyle S_{M_2}(z) & = & \scriptstyle \frac{1+z}{z} \Upsilon^{-1}_{M_2}(z), \; \mbox{from (\ref{Eq52})} \\
 \scriptstyle S_{V_2 \frac{P_2}{\alpha_{22}} V_2^H}\left(\frac{z}{\alpha_{22}}\right)
\nonumber & =& \scriptstyle \frac{1 + \frac{z}{\alpha_{22}}}{\frac{z}{\alpha_{22}}} \Upsilon^{-1}_{V_2 \frac{P_2}{\alpha_{22}} V_2^H} \left(\frac{z}{\alpha_{22}}\right),\: \mbox{from (\ref{Eq52})} \label{EqS_M2_2}\\
& =& \scriptstyle \frac{\alpha_{22} + z}{z} \Upsilon^{-1}_{V_2 \frac{P_2}{\alpha_{22}} V_2^H}\left(\frac{z}{\alpha_{22}}\right) \label{EqS_V2P2V2h}
\end{eqnarray}
Then, plugging (\ref{EqS_M2}) and (\ref{EqS_V2P2V2h}) into
(\ref{EqG_M2}) yields
\begin{eqnarray}\label{EqUpsilon_M2}
\Upsilon^{-1}_{M_2}(z)
&=& \left(\frac{1}{1+z}\right) \Upsilon^{-1}_{V_2 \frac{P_2}{\alpha_{22}} V_2^H}\left(\frac{z}{\alpha_{22}}\right)
\end{eqnarray}
Now, using the relation (\ref{Eq53a}) between the
$\Upsilon$-transform and the Stieltjes transform, we write
\begin{equation}\label{}
G_{M_2}(z) = \left(\frac{-1}{z}\right) \left(\Upsilon_{M_2}\left(\frac{1}{z}\right) + 1 \right),
\end{equation}
and from (\ref{EqG_M}), we obtain
\begin{eqnarray}\label{EqGM}
\textstyle G_{M}(z) = \textstyle \left( \frac{-1}{z - g(G_M(z))} \right) \left(\Upsilon_{M_2}\left(\frac{1}{z - g(G_M(z))}  \right) + 1 \right).
\end{eqnarray}
We handle  (\ref{EqGM}) to obtain $G_{M}(z)$ as a function of
$\Upsilon_{V_2 P_2 V_2^H}(z)$:
\begin{eqnarray}
\label{EqG_M_FPE} \scriptstyle \Upsilon_{M_2}\left(\frac{1}{z - g(G_M(z))}\right) &=& \scriptstyle - 1 -(\: z - g(G_M(z)) \: ) \: G_{M}(z)\\
\nonumber \scriptstyle \frac{1}{z - g(G_M(z))} &=& \scriptstyle \Upsilon_{M_2}^{-1} \left( - 1 - (\: z - g(G_M(z))\:) \: G_{M}(z)  \right)\\
\nonumber \scriptstyle \frac{1}{z - g(G_M(z))} &=& \scriptstyle \frac{-1}{ (\: z - g(G_M(z))  \:) \: G_{M}(z)} \\
\nonumber & & \scriptstyle \Upsilon_{V_2 \frac{P_2}{\alpha_{22}} V_2^H}^{-1} \left( -\frac{1+ (\: z - g(G_M(z))  \:) \:G_{M}(z)}{\alpha_{22}} \right)\\
\nonumber - G_M(z) &=& \scriptstyle \Upsilon_{V_2 \frac{P_2}{\alpha_{22}} V_2^H}^{-1} \left( -\frac{1+ (\: z - g(G_M(z))  \:) \:G_{M}(z)}{\alpha_{22}} \right) \\
\nonumber \scriptstyle \Upsilon_{V_2 \frac{P_2}{\alpha_{22}} V_2^H} \left( - G_M(z)\right) &=& \scriptstyle -\frac{1+ (\: z - g(G_M(z))  \:) \:G_{M}(z)}{\alpha_{22}}\\
\nonumber \scriptstyle G_{M}(z) &=& \scriptstyle \left( - \frac{1}{z - g(G_M(z))} \right) \\
\nonumber 										  & & \scriptstyle \left( 1+ \alpha_{22} \Upsilon_{V_2 \frac{P_2}{\alpha_{22}} V_2^H} \left( -
															       G_M(z) \right) \right).
\end{eqnarray}
\noindent
From the definition of the $\Upsilon$-transform (\ref{Eq53}), it follows
that
\begin{eqnarray}
\nonumber \scriptstyle \alpha_{22} \Upsilon_{V_2 \frac{P_2}{\alpha_{22}} V_2^H} \left(- G_M(z)\right)
 &=& \scriptstyle \alpha_{22}\textstyle\int \scriptstyle\frac{-G_M(z) \lambda }{ 1+ G_M(z) \lambda}  \d F_{V_2 \frac{P_2}{\alpha_{22}} V_2^H}(\lambda)\\
\nonumber &=& \textstyle\int \scriptstyle\frac{- \alpha_{22}G_M(z) \lambda }{ 1+ G_M(z) \lambda}  \alpha_{22}f_{V_2 P_2 V_2^H}(\alpha_{22}\lambda)\d \lambda \\
&=& \textstyle\int \scriptstyle\frac{-  G_M(z) t}{ 1+ G_M(z) \frac{t}{\alpha_{22}}} f_{V_2 P_2 V_2^H}(t) \mathrm{d}t\label{eq:UpsV2P2V2}
\end{eqnarray}

\noindent
Using (\ref{eq:UpsV2P2V2}) in (\ref{EqG_M_FPE}), we have
\begin{equation}\label{eq:GM}
\scriptstyle G_{M}(z)
= \left( - \frac{1}{z - g(G_M(z))} \right)  \left( 1 - G_M(z) \: h(G_M(z)) \: \right)
\end{equation}
with the function $h(u)$ defined as follows
\begin{eqnarray}
\nonumber h(u) &\triangleq& \displaystyle\int\frac{t}{ 1+ \frac{u}{\alpha_{22}} t}  \d F_{V_2 P_2 V_2^H}(t) = \mathds{E}\left[ \frac{p_2}{1+ \frac{1}{\alpha_{22}} p_2 u} \right]
\end{eqnarray}
where the random variable $p_2$ follows the distribution $F_{V_2P_2V_2^H}$.

\noindent
Factorizing $G_M(z)$  in (\ref{eq:GM}) finally yields
\begin{equation}\label{Eq_GM_Final}
G_M(z) = \frac{-1}{z-g(G_M(z))-h(G_M(z))}
\end{equation}

\noindent
Expression (\ref{Eq_GM_Final}) is a fixed point equation with unique
solution when $z \in \mathds{R}_{-}$ \cite{Silverstein-95a}.

\noindent
\textbf{Step 4: Integrate $\frac{\partial \bar{R}_2(\bs{P}_2,\sigma^2_2)}{\partial\sigma^2_2}$ to obtain  $\bar{R}_2(\bs{P}_2,\sigma^2_2)$ in the RLNA.}

\noindent
From (\ref{Eq20}), we have that
\begin{eqnarray}\label{EqDerivative}
\scriptstyle \frac{\partial}{\partial\sigma^2_2} \bar{R}_{2,\infty}(\bs{P}_2, \sigma^2_2) = \frac{1}{\ln2} \left( G_M\left(-\sigma^2_2\right) - G_{M_1}\left(-\sigma^2_2\right) \right).
\end{eqnarray}
Moreover, it is know that if $\sigma^2_2 \rightarrow \infty$ no
reliable communication is possible and thus,
$\bar{R}_{2,\infty} = 0$. Hence, the asymptotic
rate of the opportunistic link can be obtained by integrating
expression (\ref{EqDerivative})
\begin{equation}\label{EqOpportunisticCapacity}
\begin{split}
\bar{R}_{2,\infty}
&=\frac{-1}{\ln2}\displaystyle\int_{\sigma^2_2}^{\infty} \left( G_M\left(-z\right) - G_{M_1}\left(-z\right)\right) \mathrm{d}z,
\end{split}
\end{equation}
which ends the proof.
\bibliographystyle{IEEEtran}
\bibliography{IEEEabrv,IA-Transactions}
%


\end{document}